\documentclass[11pt,a4paper]{article}
\usepackage{jheppub}
\usepackage{epsfig}
\usepackage{amsmath}
\usepackage{latexsym}
\usepackage{amssymb}
\usepackage{dsfont}
\usepackage{color} 
\usepackage{multirow}
\usepackage{enumitem}

\newcommand{\ud}{\mathrm{d}}

\newcommand{\uTr}{\mathrm{Tr}}

\newlength\savedwidth

\title{Structure analysis of the generalized correlator of quark and gluon for a spin-$1/2$ target}

\author[a]{C. Lorc\'e,}
\author[b]{B. Pasquini}

\affiliation[a]{IPNO, Universit\'e Paris-Sud, CNRS/IN2P3, 91406 Orsay, France\\
and LPT, Universit\'e Paris-Sud, CNRS, 91405 Orsay, France}
\affiliation[b]{Dipartimento di Fisica, Universit\`a degli Studi di Pavia, \\and INFN, Sezione di Pavia, I-27100 Pavia, Italy}
\emailAdd{lorce@ipno.in2p3.fr}
\emailAdd{pasquini@pv.infn.it}

\abstract
{
We analyze the structure of generalized off-diagonal and transverse-momentum dependent quark-quark and gluon-gluon correlators for a spin-$1/2$ hadron. Using the light-front formalism, we provide a parametrization in terms of the parton generalized transverse-momentum dependent distributions that emphasizes the multipole structure of the correlator. The results for the quark-quark correlation functions are consistent with an alternative parametrization given in terms of Lorentz covariant structures. The parametrization for the gluon-gluon generalized correlator is presented for  the first time and allows one to introduce new correlation functions which can be relevant for phenomenological applications.
}


\begin{document}
\maketitle
\flushbottom

\section{Introduction}
\label{section-1}

Both the generalized parton distributions (GPDs) \cite{Mueller:1998fv,Ji:1996ek,Radyushkin:1996nd,Goeke:2001tz,Diehl:2003ny,Belitsky:2005qn,Boffi:2007yc}, appearing in the description of hard exclusive reactions like deeply virtual Compton scattering, and the transverse-momentum dependent parton distributions (TMDs) \cite{Mulders:1995dh,Barone:2001sp,Bacchetta:2006tn,D'Alesio:2007jt}, appearing in the description of semi-inclusive reactions like semi-inclusive deep inelastic scattering and Drell-Yan process, have been intensively studied in the last two decades. These distributions provide us with essential information about the distribution and the orbital motion of partons inside hadrons, and allow us to draw three-dimensional pictures of the nucleon, either in mixed position-momentum space or in pure momentum space \cite{Boer:2011fh}. 

Despite numerous suggestions in the literature \cite{Burkardt:2002ks,Burkardt:2003uw,Burkardt:2003je,Diehl:2005jf,Burkardt:2005hp,Lu:2006kt,Meissner:2007rx}, no nontrivial model-independent relations between GPDs and TMDs have been found \cite{Meissner:2008ay,Meissner:2009ww}. However, both the GPDs and the TMDs appear to be two different limits of more general correlation functions called generalized TMDs\footnote{Such functions are also known under the name of unintegrated GPDs.} (GTMDs) \cite{Meissner:2008ay,Meissner:2009ww} which can show up in the description of hard QCD processes \cite{Collins:2007ph,Rogers:2008jk}. They depend on the 3-momentum of the partons and, in addition, contain information on the momentum transfer to the hadron. The quark GTMDs typically appear at subleading twist and in situations where the standard collinear factorization cannot be applied, see \emph{e.g.} refs.~\cite{Vanderhaeghen:1999xj,Diehl:2007hd,Goloskokov:2007nt}. On the other hand, gluon GTMDs have been extensively used in the description of high-energy processes like \emph{e.g.} diffractive vector meson production \cite{Martin:1999wb} and Higgs production at the Tevatron and the LHC \cite{Khoze:2000cy,Albrow:2008pn,Martin:2009ku} using the $k_T$-factorization framework. In ref.~\cite{Martin:2001ms} is also suggested an approximate method for constraining the unpolarized gluon GTMD. The GTMDs have  a direct connection with Wigner distributions of the parton-hadron system~\cite{Ji:2003ak,Belitsky:2003nz,Belitsky:2005qn,Lorce:2011dv} which represent the quantum-mechanical analogues of the classical phase-space distributions and have recently been discussed to access the orbital angular momentum structure of partons in  hadrons~\cite{Lorce:2011kd,Lorce:2011ni,Hatta:2011ku,Ji:2012sj}. 

The parametrization of the generalized (off-diagonal) quark-quark correlation functions for a spin-$0$ and spin-$1/2$ hadron has been given for the first time in refs.~\cite{Meissner:2008ay,Meissner:2009ww}. Here, we want to extend this study to the generalized gluon-gluon correlator, proposing  a convenient formalism which allows us to discuss in a unified framework also the quark-quark correlator. Such a formalism is based on the light-front quantization and on the analysis of the multipole pattern given by the parton operators entering the two-parton generalized correlators at different twists. We first identify the spin-flip number of each parton operator, defined in terms of  the helicity and orbital angular momentum transferred to the parton. To each spin-flip number we can then associate a well-defined multipole structure that can be represented in terms of the four-vectors at our disposal, multiplied by Lorentz scalar functions representing the parton GTMDs.

The various step of this derivation are presented as follows. In the next section we introduce the definition of the two-parton generalized correlator. In section~\ref{section-3}, we describe the derivation of  the parametrization of the generalized correlators in terms of GTMDs. In particular, we discuss the angular momentum structure and multipole pattern of the two-parton correlators at different twists. Taking into account also the constraints of discrete symmetries and hermiticity, we obtain a basis to parametrize both the quark-quark and the gluon-gluon  correlation functions. The results in the gluon sector are given here for the first time, while  the parametrization in the quark sector is alternative, but equivalent, to that one given in terms of Lorentz covariant structures in ref.~\cite{Meissner:2009ww}. The relations between these two sets of quark GTMDs are given in the appendix. At leading twist, we also present the results for the light-front helicity amplitudes, discussing the physical interpretation of the twist-2 GTMDs in terms of nucleon and parton polarizations. In section~\ref{section-4}, we discuss the TMD limit and the GPD limit of the GTMDs, and provide the dictionary to relate them with other existing  parametrizations of the gluon and quark distribution functions. In the last section we draw our conclusions.

\section{Generalized parton correlators}
\label{section-2}

\begin{figure}[ht]
\begin{center}
\epsfig{file=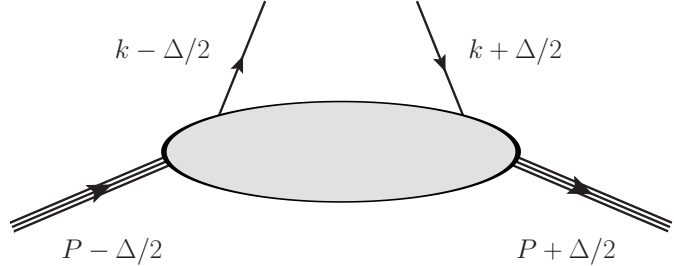,  width=0.6\columnwidth}
\end{center}
\caption{\footnotesize{Kinematics for the fully-unintegrated two-parton correlator in a symmetric frame.}}
\label{fig1}
\end{figure}

The maximum amount of information on the parton distributions inside the nucleon is contained in the fully-unintegrated two-parton correlator $W$ for a spin-$1/2$ hadron. The general quark-quark correlator is defined as\footnote{Note that these are just the naive definitions. Complications associated with \emph{e.g.} renormalization, rapidity divergences and soft factors are not addressed here as they do not affect the parametrization.}~\cite{Ji:2003ak,Belitsky:2003nz,Belitsky:2005qn,Meissner:2009ww}
\begin{equation}\label{gencorrq}
W^{[\Gamma]}_{\Lambda'\Lambda}(P,k,\Delta,N;\eta)=\frac{1}{2}\int\frac{\ud^4z}{(2\pi)^4}\,e^{ik\cdot z}\,\langle p',\Lambda'|\overline\psi(-\tfrac{z}{2})\Gamma\,\mathcal W\,\psi(\tfrac{z}{2})|p,\Lambda\rangle,
\end{equation}
and the general gluon-gluon correlator can be defined likewise
\begin{equation}\label{gencorrg}
W^{\mu\nu;\rho\sigma}_{\Lambda'\Lambda}(P,k,\Delta,N;\eta,\eta')=\frac{1}{k\cdot n}\int\frac{\ud^4z}{(2\pi)^4}\,e^{ik\cdot z}\,\langle p',\Lambda'|2\uTr\!\left[G^{\mu\nu}(-\tfrac{z}{2})\,\mathcal W\,G^{\rho\sigma}(\tfrac{z}{2})\,\mathcal W'\right]|p,\Lambda\rangle.
\end{equation}
These correlators are functions of the initial (final) hadron light-front helicity $\Lambda$ ($\Lambda'$), the average hadron four-momentum $P=(p'+p)/2$, the average parton four-momentum $k$, and the four-momentum transfer to the hadron $\Delta=p'-p$. The superscript $\Gamma$ in eq.~\eqref{gencorrq}
stands for any element of the basis $\{\mathds 1,\gamma_5,\gamma^\mu,\gamma^\mu\gamma_5,i\sigma^{\mu\nu}\gamma_5\}$ in Dirac space. The Wilson lines $\mathcal W\equiv\mathcal W(-\tfrac{z}{2},\tfrac{z}{2}|\eta n)$ and $\mathcal W'\equiv\mathcal W(\tfrac{z}{2},-\tfrac{z}{2}|\eta' n)$ ensure the color gauge invariance of the correlators~\cite{Pijlman:2006vm}, connecting the points $-\tfrac{z}{2}$ and $\tfrac{z}{2}$ \emph{via} the intermediary points $-\tfrac{z}{2}+\eta{\phantom{(}'\!\!\!^{(\phantom{'})}}\infty\cdot n$ and $\tfrac{z}{2}+\eta{\phantom{(}'\!\!\!^{(\phantom{'})}}\infty\cdot n$ by straight lines\footnote{More complicated Wilson lines can also be relevant depending on the process, see refs. \cite{Buffing:2011mj,Buffing:2012sz,Buffing:2013kca}.}, where $n$ is a lightlike vector $n^2=0$. Since any rescaled four-vector $\alpha n$ with some positive parameter $\alpha$ could be used to specify the Wilson lines, the correlators actually depend on the four-vector
\begin{equation}
N=\frac{P^2\,n}{P\cdot n}.
\end{equation}
The parameters $\eta{\phantom{(}'\!\!\!^{(\phantom{'})}}$ indicate whether the Wilson lines are future-pointing ($\eta{\phantom{(}'\!\!\!^{(\phantom{'})}}=+1$) or past-pointing ($\eta{\phantom{(}'\!\!\!^{(\phantom{'})}}=-1$). For convenience, we choose the spatial axes such that $\vec n\propto\vec e_z$ and work in a symmetric frame, see figure~\ref{fig1}.

The two-parton correlators defining TMDs, GPDs, PDFs, FFs and charges are obtained by considering specific limits or projections of  eqs.~\eqref{gencorrq} and \eqref{gencorrg}. These correlators have in common the fact that the parton fields are taken at the same light-front time $z^+=0$. We then focus our attention on the $k^-$-integrated version of eqs.~\eqref{gencorrq} and \eqref{gencorrg}
\begin{align}
W^{[\Gamma]}_{\Lambda'\Lambda}(P,x,\vec k_T,\Delta,N;\eta)&=\int\ud k^-\,W^{[\Gamma]}_{\Lambda'\Lambda}(P,k,\Delta,N;\eta)\nonumber\\\label{qGTMD}
&\hspace{-3.5cm}=\frac{1}{2}\int\frac{\ud z^-\,\ud^2z_T}{(2\pi)^3}\,e^{ixP^+z^--i\vec k_T\cdot\vec z_T}\,\langle p',\Lambda'|\overline\psi(-\tfrac{z}{2})\Gamma\,\mathcal W\,\psi(\tfrac{z}{2})|p,\Lambda\rangle\Big|_{z^+=0},\\
W^{\mu\nu;\rho\sigma}_{\Lambda'\Lambda}(P,x,\vec k_T,\Delta,N;\eta,\eta')&=\int\ud k^-\,W^{\mu\nu;\rho\sigma}_{\Lambda'\Lambda}(P,k,\Delta,N;\eta,\eta')\nonumber\\\label{gGTMD}
&\hspace{-3.5cm}=\frac{1}{xP^+}\int\frac{\ud z^-\,\ud^2z_T}{(2\pi)^3}\,e^{ixP^+z^--i\vec k_T\cdot\vec z_T}\,\langle p',\Lambda'|2\uTr\!\left[G^{\mu\nu}(-\tfrac{z}{2})\,\mathcal W\,G^{\rho\sigma}(\tfrac{z}{2})\,\mathcal W'\right]|p,\Lambda\rangle\Big|_{z^+=0},
\end{align}
where we used for a generic four-vector $a^\mu=[a^+,a^-,\vec a_T]$ the light-front components $a^\pm=(a^0\pm a^3)/\sqrt{2}$ and the transverse components $\vec a_T=(a^1,a^2)$, and where $x=k^+/P^+$ is the fraction of average longitudinal momentum and $\vec k_T$ is the average transverse momentum of the parton. These correlators are parametrized in terms of the so-called GTMDs, which can be considered as the \emph{mother distributions} of GPDs and TMDs. A complete parametrization of the quark-quark correlator \eqref{qGTMD} in terms of GTMDs has been given in ref.~\cite{Meissner:2009ww}. In the present work, we give for the first time a complete parametrization of the gluon-gluon correlator \eqref{gGTMD}, and provide the dictionary between the corresponding daughter functions (GPDs, TMDs, PDFs) and other partial parametrizations given in the literature. Moreover, we present an alternative (but equivalent) parametrization of the quark-quark correlator \eqref{qGTMD} which emphasizes better the underlying multipole pattern.

\section{Parametrization}
\label{section-3}
The correlators \eqref{qGTMD} and \eqref{gGTMD} can generally be written as
\begin{align}
W^O_{\Lambda'\Lambda}
&=\int\frac{\ud z^-\,\ud^2z_T}{(2\pi)^3}\,e^{ixP^+z^--i\vec k_T\cdot\vec z_T}\,\langle p',\Lambda'|{\cal O}(z)|p,\Lambda\rangle\Big|_{z^+=0}
\nonumber\\
&=\overline u(p',\Lambda')M^Ou(p,\Lambda),\label{gencorr}
\end{align}
where ${\cal O}(z)$ stands for the relevant quark or gluon operator, and $M^O$ is a matrix in Dirac space, with $O=[\Gamma]$ in the quark sector and $O=\mu\nu;\rho\sigma$ in the gluon sector. A general, model-independent parametrization of these objects is obtained by giving an explicit form of $M^O$ in terms of the four-vectors at our disposal ($P,k,\Delta,N$), of the Dirac matrices $(\mathds 1,\gamma_5,\gamma^\mu,\cdots)$, of the invariant tensors $g_{\mu\nu}$ and $\epsilon_{\mu\nu\rho\sigma}$, and of Lorentz scalar functions $X(x,\xi,\vec k_T^2,\vec k_T\cdot\vec\Delta_T,\vec\Delta_T^2;\eta_i)$ where, for convenience, we denoted the set of all parameters $\eta$ simply by $\eta_i$. 

Traditionally, one writes down all the possible structures compatible with the Lorentz covariance, the discrete  symmetry and the hermiticity constraints. All the allowed structures are usually not independent. Using on-shell relations like \emph{e.g.} the Gordon identities, one can eventually extract an independent subset. Such an independent subset can be thought of as a \emph{basis} for the parametrization of the correlators. Note however that because of the on-shell identities, one has a certain freedom in choosing the actual basis. Most of the time, the basis with the simplest structures is chosen. However, such a choice will generally not display the underlying twist and multipole patterns. As a result, the corresponding Lorentz scalar functions have often no simple physical interpretation.

Alternatively, one can use the light-front formalism. It has the advantage of unravelling the underlying twist and multipole patterns. Another advantage is that it is also much easier in practice, especially when there are many four-vectors at our disposal. The two methods are of course equivalent. They lead at the end to the same number of independent structures and can be translated into each other.

\subsection{Angular momentum and multipole pattern}

The quark spinors $\psi(k,\lambda)$ and gluon polarization four-vectors $\varepsilon^\mu(k,\lambda)$ have definite light-front helicity $\lambda$ corresponding to the eigenvalue of $\hat J_z=\hat S_z+\hat L_z$, where $\hat S_z$ is the standard spin operator and $\hat L_ z$ is the orbital angular momentum (OAM) operator given in momentum space by
\begin{equation}
\begin{split}
\hat L_z&=-i\left(\vec k_T\times\vec\nabla_{k_T}\right)_z\\
&=k_R\,\frac{\partial}{\partial k_R}-k_L\,\frac{\partial}{\partial k_L}.
\end{split}
\end{equation}
When discussing the angular momentum along the $z$ direction, it is convenient to use the polar combinations $a_{R,L}=a^1\pm ia^2$ for the transverse indices.

It turns out to be particularly convenient to work with a complete set of partonic operators having a well-defined spin-flip number defined as $\Delta S_z=\lambda'-\lambda+\Delta L_z$, where $\lambda$ ($\lambda'$) is the initial (final) parton light-front helicity and $\Delta L_z$ is the eigenvalue of the operator
\begin{equation}\label{OAMop}
\begin{split}
\Delta\hat L_z&=\hat L_z-\hat L'_z\\
&=k_R\,\frac{\partial}{\partial k_R}-k_L\,\frac{\partial}{\partial k_L}+\Delta_R\,\frac{\partial}{\partial\Delta_R}-\Delta_L\,\frac{\partial}{\partial\Delta_L},
\end{split}
\end{equation}
where $k=(k_f+k_i)/2$ and $\Delta=k_f-k_i$ with $k_i$ ($k_f$) the initial (final) parton momentum. For example, one can easily see that the generic structure $k_R^{m_1}k_L^{m_2}\Delta_R^{m_3}\Delta_L^{m_4}$ carries $m_1-m_2+m_3-m_4$ units of OAM. For the quark operators, we have
\begin{align}
\Delta S_z&=0&&S,\,P,\,V^\pm,\,A^\pm,\,T^{+-},\,\tfrac{1}{2}\,T^{LR},\label{opin}\\
\Delta S_z&=+1&&V^R,\,A^R,\,T^{R\pm},\\
\Delta S_z&=-1&&V^L,\,A^L,\,T^{L\pm},
\end{align}
where the scalar, pseudoscalar, vector, axial-vector and pseudotensor quark bilinears are respectively given by
\begin{align}
S&=\overline\psi\psi,\\
P&=\overline\psi\gamma_5\psi,\\
V^\mu&=\overline\psi\gamma^\mu\psi,\\
A^\mu&=\overline\psi\gamma^\mu\gamma_5\psi,\\
T^{\mu\nu}&=\overline\psi i\sigma^{\mu\nu}\gamma_5\psi.
\end{align}
For the gluon operators, we have
\begin{align}
\Delta S_z&=0&\begin{split}&\delta^{ij}_T\Gamma^{\pm i;\pm j},\,-i\epsilon^{ij}_T\Gamma^{\pm i;\pm j},\,\Gamma^{+-,+-},\,\delta^{ij}_T\Gamma^{\{+i;-j\}},\,\delta^{ij}_T\Gamma^{[+i;-j]},\\
&-i\epsilon^{ij}_T\Gamma^{\{+i;-j\}},\,-i\epsilon^{ij}_T\Gamma^{[+i;-j]},\,\tfrac{1}{4}\,\Gamma^{LR;LR},\,\tfrac{1}{2}\,\Gamma^{\{+-;LR\}},\,\tfrac{1}{2}\,\Gamma^{[+-;LR]},\end{split}\\
\Delta S_z&=+1&&\Gamma^{\{+-;\pm R\}},\,\Gamma^{[+-;\pm R]},\,\tfrac{1}{2}\,\Gamma^{\{LR;\pm R\}},\,\tfrac{1}{2}\,\Gamma^{[LR;\pm R]},\\
\Delta S_z&=-1&&\Gamma^{\{+-;\pm L\}},\,\Gamma^{[+-;\pm L]},\,\tfrac{1}{2}\,\Gamma^{\{LR;\pm L\}},\,\tfrac{1}{2}\,\Gamma^{[LR;\pm L]},\\
\Delta S_z&=+2&&\Gamma^{\pm R;\pm R},\,\Gamma^{\{+R;-R\}},\,\Gamma^{[+R;-R]},\\
\Delta S_z&=-2&&\Gamma^{\pm L;\pm L},\,\Gamma^{\{+L;-L\}},\,\Gamma^{[+L;-L]},\label{opfin}
\end{align}
where $i,j=1,2$ are transverse indices, $\epsilon^{12}_T\equiv\epsilon^{-+12}=+1$, and where we have defined
\begin{eqnarray}
\Gamma^{\mu\nu;\rho\sigma}&=&2\uTr\!\left[G^{\mu\nu}(-\tfrac{z}{2})\,\mathcal W\,G^{\rho\sigma}(\tfrac{z}{2})\,\mathcal W'\right]/xP^+,\\
\Gamma^{\{\mu\nu;\rho\sigma\}}&=&\tfrac{1}{2}(\Gamma^{\mu\nu;\rho\sigma}+\Gamma^{\rho\sigma;\mu\nu}),\\
\Gamma^{[\mu\nu;\rho\sigma]}&=&\tfrac{1}{2}(\Gamma^{\mu\nu;\rho\sigma}-\Gamma^{\rho\sigma;\mu\nu}).
\end{eqnarray}
Interestingly, the twist-2 partonic operators have $\Delta L_z=0$ leading therefore to a simple interpretation in terms of light-front helicities $\Delta S_z=\lambda'-\lambda$. For the higher-twist partonic operators, a simple interpretation does not exist since the light-front helicities are usually mixed with the OAM.  
\\
Just like the quark spinors and the gluon polarization four-vectors, the nucleon states $|p,\Lambda\rangle$ have definite light-front helicity $\Lambda$ corresponding to the eigenvalue of $\hat J_z=\hat S_z+\hat L_z$. By conservation of angular momentum, the amplitude $W^O_{\Lambda'\Lambda}$ is associated with the change of OAM $\Delta\ell_z=\Delta S_z-(\Lambda'-\Lambda)$. Since $\vec k_T$ and $\vec\Delta_T$ are the only possible transverse vectors available\footnote{By definition, $N$ does not have any transverse component. Moreover, one has always the possibility to choose a light-front frame such that $\vec P_T=\vec 0_T$. This is related to the fact that, thanks to translation invariance, a parametrization does not actually depend on $P$ apart from a trivial global factor.}, $\Delta\ell_z$ has to coincide with the eigenvalue obtained by applying the OAM operator \eqref{OAMop} to the amplitude $W^O_{\Lambda'\Lambda}$. Therefore, the general structure of the amplitude $W^O_{\Lambda'\Lambda}$ is given in terms of explicit global powers of $\vec k_T$ and $\vec\Delta_T$, accounting for the change of OAM, multiplied by a Lorentz scalar function $X(x,\xi,\vec k_T^2,\vec k_T\cdot\vec\Delta_T,\vec\Delta_T^2;\eta_i)$. Since any structure of the form $(k_L\Delta_R+k_R\Delta_L)/2=\vec k_T\cdot\vec\Delta_T$, $k_Rk_L=\vec k_T^2$ or $\Delta_R\Delta_L=\vec\Delta_T^2$ can be reabsorbed in the definition of the Lorentz scalar functions $X(x,\xi,\vec k_T^2,\vec k_T\cdot\vec\Delta_T,\vec\Delta_T^2;\eta_i)$, there can only be 2 independent structures for each value of $\Delta \ell_z$. For $\Delta\ell_z=0$, we choose $1$ and $\frac{i(\vec k_T\times\vec\Delta_T)_z}{M^2}$, while for $\Delta\ell_z=\pm m$ with $m>0$, we choose $\frac{k^m_{R(L)}}{M^m}$ and $\frac{\Delta^m_{R(L)}}{M^m}$ as the independent explicit global structures. Powers of the nucleon mass $M$ have been added such that each structure has vanishing mass dimension. As a result, each amplitude $W^O_{\Lambda'\Lambda}$ can be written in one of the following forms
\begin{align}
\Delta\ell_z&=0&S&=S_a+\frac{i(\vec k_T\times\vec\Delta_T)_z}{M^2}\,S_b,\label{struc1}\\
\Delta\ell_z&=\pm 1&P_{R(L)}&=\frac{k_{R(L)}}{M}\,P_a+\frac{\Delta_{R(L)}}{M}\,P_b,\label{struc2}\\
\Delta\ell_z&=\pm 2&D_{R(L)}&=\frac{k^2_{R(L)}}{M^2}\,D_a+\frac{\Delta^2_{R(L)}}{M^2}\,D_b,\label{struc3}\\
\Delta\ell_z&=\pm 3&F_{R(L)}&=\frac{k^3_{R(L)}}{M^3}\,F_a+\frac{\Delta^3_{R(L)}}{M^3}\,F_b,\label{struc4}\\
&\,\,\,\vdots&&\,\,\,\vdots\nonumber
\end{align}
where $a,b$ simply label the Lorentz scalar functions associated with the two independent structures for a given $\Delta\ell_z$.

\subsection{Discrete symmetry and hermiticity constraints}
\label{symmetries}

The hermiticity constraint relates amplitudes with initial and final light-front helicities interchanged, and changes the sign of the momentum transfer
\begin{equation}\label{hermiticity}
W^O_{\Lambda'\Lambda}(P,k,\Delta,N;\eta_i)=[W^{O_\mathsf H}_{\Lambda\Lambda'}(P,k,-\Delta,N;\eta_i)]^*,
\end{equation} 
where $a^*$ is the complex conjugate of $a$, and $O_\mathsf H$ is given by $[\Gamma]_{\mathsf H}=[\gamma^0\Gamma^\dag\gamma^0]$ for quarks and by $(\mu\nu;\rho\sigma)_{\mathsf H}=(\rho\sigma;\mu\nu)^*$ for gluons. For later convenience, we will use the notation $a^\Delta$ to indicate that the sign of $\Delta$ has been changed in the function $a$. 
 
For the discrete symmetries, it is convenient to use the ones adapted to the light-front coordinates \cite{Brodsky:2006ez,Soper:1972xc,Carlson:2003je}. The light-front parity changes the sign of the $a^1$ component of any four-vector $a$ and flips the light-front helicities
\begin{equation}\label{parity}
W^O_{\Lambda'\Lambda}(P,k,\Delta,N;\eta_i)=W^{O_\mathsf P}_{-\Lambda'-\Lambda}(\tilde P,\tilde k,\tilde\Delta,\tilde N;\eta_i),
\end{equation} 
where $\tilde a=[a^+,a^-, -a^1,a^2]$, \emph{i.e.} $\tilde a_{R(L)}=-a_{L(R)}$, and $O_\mathsf P$ is given by $[\Gamma]_{\mathsf P}=[(\gamma^1\gamma_5)\Gamma(\gamma^1\gamma_5)]$ for quarks and by $(\mu\nu;\rho\sigma)_{\mathsf P}=\tilde\mu\tilde\nu;\tilde\rho\tilde\sigma$ for gluons. 

Finally, under light-front time-reversal any four-momentum transforms as $q\mapsto\tilde q$, while any position four-vector transforms as $x\mapsto -\tilde x$. As a result, invariance under light-front time-reversal implies
\begin{equation}\label{TR}
W^O_{\Lambda'\Lambda}(P,k,\Delta,N;\eta_i)=[W^{O_\mathsf T}_{\Lambda'\Lambda}(\tilde P,\tilde k,\tilde\Delta,N;-\eta_i)]^*,
\end{equation}
where $O_\mathsf T$ is given by $[\Gamma]_{\mathsf T}=[(-i\gamma^1\gamma^2)\Gamma^*(-i\gamma^1\gamma^2)]$ for quarks and by $(\mu\nu;\rho\sigma)_{\mathsf T}=(\tilde\mu\tilde\nu;\tilde\rho\tilde\sigma)^*$ for gluons. In the symmetric frame one has naturally $\tilde P=P$.

The momentum arguments of the Lorentz scalar functions $X(x,\xi,\vec k_T^2,\vec k_T\cdot\vec\Delta_T,\vec\Delta_T^2;\eta_i)$ are invariant under light-front parity and time-reversal transformations. For later convenience, we then introduce the following notations:
\begin{align}
\tilde S&=S_a-\frac{i(\vec k_T\times\vec\Delta_T)_z}{M^2}\,S_b,\\
\tilde P_{R(L)}&=-\frac{k_{L(R)}}{M}\,P_a-\frac{\Delta_{L(R)}}{M}\,P_b,\\
\tilde D_{R(L)}&=\frac{k^{2}_{L(R)}}{M^2}\,D_a+\frac{\Delta^{2}_{L(R)}}{M^2}\,D_b,\\
\tilde F_{R(L)}&=-\frac{k^{3}_{L(R)}}{M^3}\,F_a-\frac{\Delta^{3}_{L(R)}}{M^3}\,F_b,\\
S^{*\Delta}(\Delta)&=S^*_a(-\Delta)+\frac{i(\vec k_T\times\vec\Delta_T)_z}{M^2}\,S^*_b(-\Delta),\\
P^{*\Delta}_{R(L)}(\Delta)&=\frac{k_{L(R)}}{M}\,P^*_a(-\Delta)-\frac{\Delta_{L(R)}}{M}\,P^*_b(-\Delta),\\
D^{*\Delta}_{R(L)}(\Delta)&=\frac{k^{2}_{L(R)}}{M^2}\,D^*_a(-\Delta)+\frac{\Delta^{2}_{L(R)}}{M^2}\,D^*_b(-\Delta),\\
F^{*\Delta}_{R(L)}(\Delta)&=\frac{k^{3}_{L(R)}}{M^3}\,D^*_a(-\Delta)-\frac{\Delta^{3}_{L(R)}}{M^3}\,D^*_b(-\Delta).
\end{align}

To each partonic operator in eqs. \eqref{opin}-\eqref{opfin}, we associate $c_{\mathsf H}$, $c_{\mathsf P}$ and $c_{\mathsf T}$ coefficients determining their properties under hermiticity, light-front parity and light-front time-reversal transformation, respectively
\begin{align}
\mathcal O_{\mathsf H}&=c_{\mathsf H}\,\mathcal O\big|_{R(L)\mapsto L(R)},\\
\mathcal O_{\mathsf P}&=c_{\mathsf P}\,\mathcal O\big|_{R(L)\mapsto -L(-R)},\label{parity-quark}\\
\mathcal O_{\mathsf T}&=c_{\mathsf T}\,\mathcal O\big|_{R(L)\mapsto -R(-L)},
\end{align}
where the replacement rule affects only the uncontracted transverse indices. An explicit pair of indices $\mathcal O^{LR}$ has to be considered as contracted since it can be rewritten in terms of $\delta_T^{ij}\mathcal O^{ij}$ and $-i\epsilon_T^{ij}\mathcal O^{ij}$. We chose the factors of $i$ in the partonic operators \eqref{opin}-\eqref{opfin} such that $c_{\mathsf T}=+1$. For the quark operators, we have
\begin{align}
c_{\mathsf H}=+1,\,c_{\mathsf P}&=+1&&S,\,V^\pm,\,V^{R(L)},\\
c_{\mathsf H}=+1,\,c_{\mathsf P}&=-1&&A^\pm,\,A^{R(L)},\,T^{+-},\,T^{R(L)\pm},\\
c_{\mathsf H}=-1,\,c_{\mathsf P}&=+1&&\tfrac{1}{2}\,T^{LR},\\
c_{\mathsf H}=-1,\,c_{\mathsf P}&=-1&&P,
\end{align}
and for the gluon operators, we have
\begin{align}
c_{\mathsf H}=+1,\,c_{\mathsf P}&=+1&\begin{split}&\delta^{ij}_T\Gamma^{\pm i;\pm j},\,\Gamma^{+-,+-},\,\delta^{ij}_T\Gamma^{\{+i;-j\}},\,\tfrac{1}{4}\,\Gamma^{LR;LR},\\
&\Gamma^{\{+-;\pm R(L)\}},\,\Gamma^{\pm R(L);\pm R(L)},\,\Gamma^{\{+R(L);-R(L)\}},\end{split}\\
c_{\mathsf H}=+1,\,c_{\mathsf P}&=-1&&-i\epsilon^{ij}_T\Gamma^{\pm i;\pm j},\,-i\epsilon^{ij}_T\Gamma^{[+i;-j]},\,\tfrac{1}{2}\,\Gamma^{[+-;LR]},\,\tfrac{1}{2}\,\Gamma^{[LR;\pm R(L)]},\\
c_{\mathsf H}=-1,\,c_{\mathsf P}&=+1&&\delta^{ij}_T\Gamma^{[+i;-j]},\,\Gamma^{[+-;\pm R(L)]},\,\Gamma^{[+R(L);-R(L)]},\\
c_{\mathsf H}=-1,\,c_{\mathsf P}&=-1&&-i\epsilon^{ij}_T\Gamma^{\{+i;-j\}},\,\tfrac{1}{2}\,\Gamma^{\{+-;LR\}},\,\tfrac{1}{2}\,\Gamma^{\{LR;\pm R(L)\}}.
\end{align}

\subsection{Quark and gluon GTMDs}
  
For a given partonic operator $\cal O$, the amplitude $W^O_{\Lambda'\Lambda}$ can conveniently be represented as a $2\times 2$ matrix in the proton light-front helicity basis. The amplitudes with $\Delta S_z=0$ and parity $c_{\mathsf P}=\pm 1$ have the following generic structure
\begin{equation}\label{Mat0}
A^{0,\, c_{\mathsf P}}_{\Lambda'\Lambda}=
\begin{pmatrix}
S&c_{\mathsf P}\tilde P_R\\
P_R&c_{\mathsf P}\tilde S
\end{pmatrix},
\end{equation}
where the row entries correspond to $\Lambda'=\tfrac{1}{2},-\tfrac{1}{2}$ and the column entries are likewise $\Lambda=\tfrac{1}{2},-\tfrac{1}{2}$. Furthermore, the hermiticity constraint imposes  the following relations  
\begin{equation}\label{cond1}
S=c_{\mathsf H}S^{*\Delta},\qquad P_R=c_{\mathsf H}c_{\mathsf P}\tilde P^{*\Delta}_R.
\end{equation}
Similarly, we have the following generic structure for $\Delta S_z=\pm 1$ 
\begin{equation}\label{Mat1}
 A^{+1,\, c_{\mathsf P}}_{\Lambda'\Lambda}=\begin{pmatrix}
 P_R+P'_R& S\\
D_R&P_R-P'_R
\end{pmatrix},
\qquad
A^{-1,\, c_{\mathsf P}}_{\Lambda'\Lambda}=-c_{\mathsf P} 
\begin{pmatrix}
\tilde P_R-\tilde P'_R&\tilde D_R\\
\tilde S&\tilde P_R+\tilde P'_R
\end{pmatrix},
\end{equation}
where the hermiticity constraint imposes
\begin{equation}\label{cond2}
S=-c_{\mathsf H}c_{\mathsf P}\tilde S^{*\Delta}, 
\qquad P_R=-c_{\mathsf H}c_{\mathsf P}\tilde P^{*\Delta}_R, 
\qquad P'_R=c_{\mathsf H}c_{\mathsf P}\tilde P'^{*\Delta}_R,\qquad D_R=-c_{\mathsf H}c_{\mathsf P}\tilde D^{*\Delta}_R.
\end{equation}
Finally, we have the following generic structure for $\Delta S_z=\pm 2$ 
\begin{equation}\label{Mat2}
A^{+2,c_{\mathsf P}}_{\Lambda'\Lambda}=
\begin{pmatrix}
 D_R+D'_R&P_R\\
F_R& D_R-D'_R
\end{pmatrix},\qquad
A^{-2,c_{\mathsf P}}_{\Lambda'\Lambda}=c_{\mathsf P}
\begin{pmatrix}
\tilde D_R-\tilde D'_R&\tilde F_R\\
\tilde P_R&\tilde D_R+\tilde D'_R
\end{pmatrix},
\end{equation}
where the hermiticity constraint imposes
\begin{equation}\label{cond3}
P_R=c_{\mathsf H}c_{\mathsf P}\tilde P^{*\Delta}_R, \qquad D_R=c_{\mathsf H}c_{\mathsf P}\tilde D^{*\Delta}_R,  \qquad D'_R=-c_{\mathsf H}c_{\mathsf P}\tilde D'^{*\Delta}_R, \qquad F_R=c_{\mathsf H}c_{\mathsf P}\tilde F^{*\Delta}_R.
\end{equation}

The $2\times 2$ matrices in eqs. \eqref{Mat0}, \eqref{Mat1} and \eqref{Mat2} can be expressed in the more conventional bilinear form
\begin{equation}
\label{P0}
A^{\Delta S_z,c_{\mathsf P}}_{\Lambda'\Lambda}=\frac{\overline u(p',\Lambda')M^{\Delta S_z,c_{\mathsf P}}u(p,\Lambda)}{2P^+\sqrt{1-\xi^2}},
\end{equation}
where $M^{\Delta S_z}$ is a Dirac matrix. The general structure of these matrices can be written in the following form:
{\allowdisplaybreaks
\begin{align}
\label{P1}
M^{0,+}&=\left(\frac{M}{P^+}\right)^{t-1}\left[\gamma^+\left(S^{0,+}_{t,ia}+\gamma_5\,\frac{i\epsilon^{k_T\Delta_T}_T}{M^2}\,S^{0,+}_{t,ib}\right)+i\sigma^{j+}\left(\frac{k^j_T}{M}\,P^{0,+}_{t,ia}+\frac{\Delta^j_T}{M}\,P^{0,+}_{t,ib}\right)\right],\\
\label{P2}
M^{0,-}&=\left(\frac{M}{P^+}\right)^{t-1}\left[\gamma^+\gamma_5\left(S^{0,-}_{t,ia}+\gamma_5\,\frac{i\epsilon^{k_T\Delta_T}_T}{M^2}\,S^{0,-}_{t,ib}\right)+i\sigma^{j+}\gamma_5\left(\frac{k^j_T}{M}\,P^{0,-}_{t,ia}+\frac{\Delta^j_T}{M}\,P^{0,-}_{t,ib}\right)\right],\\
\label{P3}
M^{+1,+}&=\left(\frac{M}{P^+}\right)^{t-1}\left[\gamma^+\left(\frac{k_R}{M}\,P^{1,+}_{t,ia}+\frac{\Delta_R}{M}\,P^{1,+}_{t,ib}\right)+\gamma^+\gamma_5\,i\epsilon_{T}^{Rj}\left(\frac{k^j_T}{M}\,P'^{1,+}_{t,ia}+\frac{\Delta^j_T}{M}\,P'^{1,+}_{t,ib}\right)
\right.\nonumber\\
&\left.\quad-\frac{i\sigma^{R+}}{2}\left(S^{1,+}_{t,ia}-\gamma_5\,\frac{i\epsilon^{k_T\Delta_T}_T}{M^2}\,S^{1,+}_{t,ib}\right)+\frac{i\sigma^{L+}}{2}\left(\frac{k^2_R}{M^2}\,D^{1,+}_{t,ia}+\frac{\Delta^2_R}{M^2}\,D^{1,+}_{t,ib}\right)\right],\\
\label{P4}
M^{+1,-}&=\left(\frac{M}{P^+}\right)^{t-1}\left[\gamma^+\,i\epsilon^{Rj}_T\left(\frac{k^j_T}{M}\,P^{1,-}_{t,ia}+\frac{\Delta^j_T}{M}\,P^{1,-}_{t,ib}\right)+\gamma^+\gamma_5\left(\frac{k_R}{M}\,P'^{1,-}_{t,ia}+\frac{\Delta_R}{M}\,P'^{1,-}_{t,ib}\right)\right.\nonumber\\
&\left.\quad+\frac{i\sigma^{R+}\gamma_5}{2}\left(S^{1,-}_{t,ia}-\gamma_5\,\frac{i\epsilon^{k_T\Delta_T}_T}{M^2}\,S^{1,-}_{t,ib}\right)+\frac{i\sigma^{L+}\gamma_5}{2}\left(\frac{k_R^2}{M^2}\,D^{1,-}_{t,ia}+\frac{\Delta_R^2}{M^2}\,D^{1,-}_{t,ib}\right)\right],\\
M^{+2,+}&=\left(\frac{M}{P^+}\right)^{t-1}\left[\gamma^+\left(\frac{k^2_R}{M^2}\,D^{2,+}_{t,ia}
+\frac{\Delta^2_R}{M^2}\,D^{2,+}_{t,ib}\right)+\gamma^+\gamma_5\,i\epsilon_{T}^{Rj}\left(
\frac{k^j_Tk_R}{M^2}\,D'^{2,+}_{t,ia}+\frac{\Delta^j_T\Delta_R}{M^2}\,D'^{2,+}_{t,ib}\right)\right.\nonumber\\
&\left.\quad-\frac{i\sigma^{R+}}{2}\left(\frac{k_R}{M}\,P^{2,+}_{t,ia}
+\frac{\Delta_R}{M}\,P^{2,+}_{t,ib}\right)+\frac{i\sigma^{L+}}{2}\left(
 \frac{k_R^3}{M^3}\,F^{2,+}_{t,ia}+\frac{\Delta_R^3}{M^3}\,F^{2,+}_{t,ib}\right)\right],
\label{P5}
\end{align}
where $t+1$ is the twist of the partonic operator, and we used the notations $\epsilon^{ab}_T=\epsilon_T^{ij}a^{i}b^{j}$ and $\epsilon^{R(L)j}_T=\epsilon^{1j}_T\pm i\epsilon^{2j}_T$. In eqs.~\eqref{P1}-\eqref{P5}, the Lorentz scalar functions are labeled with an additional index $i$ to distinguish functions appearing at the same twist order and with the same value of $\Delta S_z$ and $c_{\mathsf P}$. The matrices with $\Delta S_z<0$ are simply obtained from eqs.~\eqref{P3}-\eqref{P5} \emph{via} the substitution $R(L)\mapsto L(R)$ and leaving the scalar functions unchanged. 

\begin{table}[t]
\begin{center}
\caption{\footnotesize{Quark operators entering the definition of the quark GTMDs  \eqref{gencorrq}, classified according to the twist order, the spin-flip $\Delta S_z$, and the value of the  $c_{\mathsf P}$ coefficient defining the properties under light-front parity transformations given in eq.~\eqref{parity-quark}. The integer $i$ in the second column corresponds to the label of the functions in eqs.~\eqref{P1}-\eqref{P5} and distinguishes functions associated to different operators with the same twist and the same values of $\Delta S_z$ and $c_{\mathsf P}$.}}\label{Qstruct}
\vspace{.3cm}
\begin{tabular}{c|c|c|c|c|c}
\hline\hline
\multirow{2}{*}{Twist}&\multirow{2}{*}{$i$}&\multicolumn{2}{|c|}{$\Delta S_z=0$}&\multicolumn{2}{c}{$\Delta S_z=+1(-1)$} \\
\cline{3-6}
&&$c_{\mathsf P}=+1$& $c_{\mathsf P}=-1$& $c_{\mathsf P}=+1$& $c_{\mathsf P}=-1$\\
\hline
2&$1$&$V^{+}$ &$A^{+}$ &---&$T^{R(L)+}$ \\
3&$1$&$S$&$T^{+-}$
& $V^{R(L)}$&$A^{R(L)}$\\
3&$2$& $\tfrac{1}{2}\,T^{LR}$&$P$&---&---\\
4&$1$&$V^{-}$ &$A^-$&---&$T^{R(L)-}$\\
\hline\hline
\end{tabular}
\end{center}
\end{table}

\begin{table}[t]
\begin{center}
\caption{\footnotesize{Same as table \ref{Qstruct} but for the gluon operators.}}\label{Gstruct}
\begin{tabular}{c|c|c|c|c|c|c}
\hline\hline
\multirow{2}{*}{Twist}&\multirow{2}{*}{$i$}&\multicolumn{2}{|c|}{$\Delta S_z=0$}&\multicolumn{2}{c|}{$\Delta S_z=+1(-1)$}&$\Delta S_z=+2(-2)$ \\
\cline{3-7}
&& $c_{\mathsf P}=+1$& $c_{\mathsf P}=-1$& $c_{\mathsf P}=+1$& $c_{\mathsf P}=-1$& $c_{\mathsf P}=+1$\\
\hline
2& $1$&$\delta^{ij}_T\Gamma^{+i;+j}$ &$-i\epsilon^{ij}_T\Gamma^{+i;+j}$ &--- &--- &$\Gamma^{+R(L);+R(L)}$ \\
3&$1$&---&---& $\Gamma^{\{+-;+R(L)\}}$&$\tfrac{1}{2}\,\Gamma^{[LR;+R(L)]}$&---\\
3&$2$&---&---& $\Gamma^{[+-;+R(L)]}$ &$\tfrac{1}{2}\,\Gamma^{\{LR;+R(L)\}}$&---\\
4&$1$&$\delta^{ij}_T\Gamma^{\{+i;-j\}}$  
&$-i\epsilon^{ij}_T\Gamma^{[+i;-j]}$&---&--- &$\Gamma^{\{+R(L);-R(L)\}}$  \\
4&$2$&$\delta^{ij}_T\Gamma^{[+i;-j]}$ &$-i\epsilon^{ij}_T\Gamma^{\{+i;-j\}}$&---&--- & $\Gamma^{[+R(L);-R(L)]}$ \\
4&$3$&$\Gamma^{+-;+-}$&$\tfrac{1}{2}\,\Gamma^{[+-;LR]}$ &--- &---&---\\
4&$4$&$\tfrac{1}{4}\,\Gamma^{LR;LR}$&$\tfrac{1}{2}\,\Gamma^{\{+-;LR\}}$&--- &---&---\\
5&$1$&---&---&$ \Gamma^{\{+-;-R(L)\}}$&$\tfrac{1}{2}\,\Gamma^{[LR;-R(L)]}$&---\\
5&$2$&---&---&$\Gamma^{[+-;-R(L)]}$& $\tfrac{1}{2}\,\Gamma^{\{LR;-R(L)\}}$ &---\\
6&$1$&$\delta^{ij}_T\Gamma^{-i;-j}$ &$-i\epsilon^{ij}_T\Gamma^{-i;-j}$ &--- &--- &$\Gamma^{-R(L);-R(L)}$\\
\hline\hline
\end{tabular}
\end{center}
\end{table}

The general parametrization of the GTMD correlators \eqref{qGTMD} and \eqref{gGTMD} is given by eqs.~\eqref{P0}-\eqref{P5} and is determined by the twist $t+1$, the spin-flip $\Delta S_z$ and the parity coefficient $c_{\mathsf P}$ of the partonic operator summarized in tables~\ref{Qstruct} and \ref{Gstruct}. The relations between the quark GTMDs in eqs.~\eqref{P1}-\eqref{P4} and the nomenclature introduced in ref.~\cite{Meissner:2009ww} are given in appendix~\ref{app:1}.

Each GTMD $X$ in this parametrization \eqref{P1}-\eqref{P5} is a complex-valued function. Light-front time-reversal and hermiticity constraints determine the behavior of these functions under a sign change of $\Delta$ or $\eta_i$. The light-front time-reversal constraint \eqref{TR} 
implies that
\begin{equation}
X^*(x,\xi, \vec k^2_T, \vec k_T\cdot\vec\Delta_T,\vec\Delta^2_T;\eta_i)=
X(x,\xi, \vec k^2_T, \vec k_T\cdot\vec\Delta_T,\vec\Delta^2_T;-\eta_i).\label{xt}
\end{equation}
It follows that the real part of the GTMDs is $\mathsf T$-even, \emph{i.e.} $\Re\text{e}X (-\eta_i)=\Re\text{e}X (\eta_i)$, while the imaginary part is $\mathsf T$-odd, \emph{i.e.} $\Im\text{m}X (-\eta_i)=-\Im\text{m}X (\eta_i)$. Finally, the hermiticity constraint \eqref{hermiticity} implies that
 \begin{equation}
X^*(x,\xi, \vec k^2_T, \vec k_T\cdot\vec\Delta_T,\vec\Delta^2_T;\eta_i) =\pm
X(x,-\xi, \vec k^2_T, -\vec k_T\cdot\vec\Delta_T,\vec\Delta^2_T;\eta_i),\label{xh}
\end{equation}
where the sign depends on the particular values of $c_{\mathsf H}$, $c_{\mathsf P}$ and $\Delta S_z$ according to eqs. \eqref{cond1}, \eqref{cond2} and \eqref{cond3}. The GTMDs can be sorted into two classes $X_+$ and $X_-$ depending on the sign in eq. \eqref{xh}. The complex-valued GTMDs can then be written as
\begin{align}
\label{x1}
X_+&=X^{ee}_++iX^{oo}_+,\\
\label{x2}
X_-&=X^{oe}_-+iX^{eo}_-,
\end{align}
where $X^{ee}_+=\Re\text{e}X_+$, $X^{oo}_+=\Im\text{m}X_+$, $X^{oe}_-=\Re\text{e}X_-$ and $X^{eo}_-=\Im\text{m}X_-$ are real-valued functions with definite symmetry under sign change of $\Delta$ (first superscript) and $\eta_i$ (second superscript). An even (or symmetric) function is labeled by $e$ and an odd (or antisymmetric) function is labeled by $o$. In the quark sector, the functions which belong to the class $X_+$ are 
$S^{0,+;q}_{t,1a}$,  
$S^{0,+;q}_{t,1b}$, 
$S^{0,-;q}_{t,1a}$,  
$S^{0,-;q}_{t,1b}$, 
$P^{0,+;q}_{t,1b}$, 
$P^{0,+;q}_{2,2a}$, 
$P^{0,-;q}_{t,1a}$, 
$P^{0,-;q}_{2,2b}$,  
$S^{1,+;q}_{2,1b}$, 
$S^{1,-;q}_{t,1a}$,  
$P^{1,+;q}_{2,1a}$, 
$P'^{1,+;q}_{2,1b}$, 
$P^{1,-;q}_{t,1b}$,  
$P'^{1,-;q}_{t,1a}$, 
$D^{1,-;q}_{t,1a}$,  
$D^{1,-;q}_{t,1b}$, where $t=1,2,3$. In the gluon sector, the functions which belong to the class $X_+$ are
$S^{0,+;g}_{t,1a}$, 
$S^{0,+;g}_{t,1b}$, 
$S^{0,+;g}_{3,3a}$, 
$S^{0,+;g}_{3,3b}$,
$S^{0,+;g}_{3,4a}$,
$S^{0,+;g}_{3,4b}$, 
$S^{0,-;g}_{t,1a}$,  
$S^{0,-;g}_{t,1b}$, 
$S^{0,-;g}_{3,3a}$,  
$S^{0,-;g}_{3,3b}$, 
$P^{0,+;g}_{t,1b}$, 
$P^{0,+;g}_{3,2a}$,
$P^{0,+;g}_{3,3b}$,  
$P^{0,+;g}_{3,4b}$,
$P^{0,-;g}_{t,1a}$, 
$P^{0,-;g}_{3,2b}$, 
$P^{0,-;g}_{3,3a}$, 
$P^{0,-;g}_{3,4b}$,
$S^{1,+;g}_{t',1b}$,   
$S^{1,+;g}_{t',2a}$,
$S^{1,-;g}_{t',1a}$,  
$S^{1,-;g}_{t',2b}$, 
$P^{1,+;g}_{t',1a}$, 
$P^{1,+;g}_{t',2b}$, 
$P'^{1,+;g}_{t',1b}$, 
$P'^{1,+;g}_{t',2a}$,   
$P^{1,-;g}_{t',1b}$,  
$P^{1,-;g}_{t',2a}$,
$P'^{1,-;g}_{t',1a}$,  
$P'^{1,-;g}_{t',2b}$, 
$D^{1,+;g}_{t',2a}$,  
$D^{1,+;g}_{t',2b}$,  
$D^{1,-;g}_{t',1a}$, 
$D^{1,-;g}_{t',1b}$,  
$P^{2,+;g}_{t,1b}$, 
$P^{2,+;g}_{3,2a}$,   
$D^{2,+;g}_{t,1a}$, 
$D^{2,+;g}_{t,1b}$, 
$D'^{2,+;g}_{3,2a}$, 
$D'^{2,+;g}_{3,2b}$,  
$F^{2,+;g}_{t,1b}$, 
$F^{2,+;g}_{3,2a}$,   
where $t=1,3,5$ and $t'=2,4$. All the other functions belong to the class $X_-$.

\subsection{Quark and gluon light-front helicity amplitudes}

For the two-parton correlators at leading twist, it is also convenient to represent them in terms of helicity amplitudes. We will restrict ourselves to the region $x>\xi$ where the GTMDs describe the emission of a parton with momentum $k_i$ and helicity $\lambda$ from the nucleon, and its reabsorption with momentum $k_f$ and helicity $\lambda'$. Any parton operator $\mathcal O$ occurring in the definition of the parton correlators ~\eqref{qGTMD} can be decomposed in the parton light-front helicity basis as follows $\mathcal O =\sum_{\lambda',\lambda}c_{\lambda'\lambda}\mathcal O_{\lambda'\lambda}$. The light-front helicity amplitudes are then defined as the matrix elements of  $\mathcal O_{\lambda'\lambda}$ in the states of definite hadron light-front helicities \cite{Diehl:2001pm}
\begin{equation}
H_{\Lambda'\lambda',\Lambda\lambda}(P,k,\Delta,N;\eta_i)=\langle p',\Lambda'|  \mathcal O_{\lambda'\lambda}(k,N;\eta_i)|p,\Lambda\rangle,
\end{equation}
and depend in general on all the four-vectors at our disposal.

At leading twist, the spin-flip $\Delta S_z$ associated with the partonic operator can be identified with the difference of light-front helicities of the parton between the final and initial states, \emph{i.e.} $\Delta S_z=\lambda'-\lambda$. Then, by conservation of the total angular momentum, the orbital angular momentum transfer to the parton is simply given by $\Delta \ell_z=(\Lambda-\lambda)-(\Lambda'-\lambda')$. As a result, to each value of the spin-flip $\Delta S_z$ one can associate at leading twist a well-defined state of polarization for the active parton \cite{Diehl:2003ny}. In the quark sector, $\tfrac{1}{2}\,V^+$ corresponds to the unpolarized quark operator,  $\tfrac{1}{2}\,A^+$ corresponds to the longitudinally polarized quark operator, and $\tfrac{1}{2}\,T^{R(L)+}$ correspond to the transversely polarized quark operators
\begin{align}
\frac{1}{2}\int\frac{\ud z^-\,\ud^2z_T}{(2\pi)^3}\,e^{ixP^+z^--i\vec k_T\cdot\vec z_T}\,V^+(z)&=\mathcal O^q_{+\tfrac{1}{2}+\tfrac{1}{2}}+\mathcal O^q_{-\tfrac{1}{2}-\tfrac{1}{2}}\equiv\mathcal O^q_U,\\
\frac{1}{2}\int\frac{\ud z^-\,\ud^2z_T}{(2\pi)^3}\,e^{ixP^+z^--i\vec k_T\cdot\vec z_T}\,A^+(z)&=\mathcal O^q_{+\tfrac{1}{2}+\tfrac{1}{2}}-\mathcal O^q_{-\tfrac{1}{2}-\tfrac{1}{2}}\equiv\mathcal O^q_L,\\
\frac{1}{2}\int\frac{\ud z^-\,\ud^2z_T}{(2\pi)^3}\,e^{ixP^+z^--i\vec k_T\cdot\vec z_T}\,T^{R+}(z)&=2\,\mathcal O^q_{+\tfrac{1}{2}-\tfrac{1}{2}}\equiv\mathcal O^q_{T_R},\\
\frac{1}{2}\int\frac{\ud z^-\,\ud^2z_T}{(2\pi)^3}\,e^{ixP^+z^--i\vec k_T\cdot\vec z_T}\,T^{L+}(z)&=2\,\mathcal O^q_{-\tfrac{1}{2}+\tfrac{1}{2}}\equiv\mathcal O^q_{T_L}.
\end{align}
Similarly, in the gluon sector, $\delta^{ij}_T\Gamma^{+i;+j}$ corresponds to the unpolarized gluon operator,  $-i\epsilon^{ij}_T\Gamma^{+i;+j}$ corresponds to the longitudinally polarized gluon operator, and $-\Gamma^{+R(L);+R(L)}$ correspond to the transversely polarized gluon operators
\begin{align}
\int\frac{\ud z^-\,\ud^2z_T}{(2\pi)^3}\,e^{ixP^+z^--i\vec k_T\cdot\vec z_T}\,\delta^{ij}_T\Gamma^{+i;+j}(z)&=\mathcal O^g_{+1+1}+\mathcal O^g_{-1-1}\equiv\mathcal O^g_U,\\
\int\frac{\ud z^-\,\ud^2z_T}{(2\pi)^3}\,e^{ixP^+z^--i\vec k_T\cdot\vec z_T}\,i\epsilon^{ij}_T\Gamma^{+i;j+}(z)&=\mathcal O^g_{+1+1}-\mathcal O^g_{-1-1}\equiv\mathcal O^g_L,\\
\int\frac{\ud z^-\,\ud^2z_T}{(2\pi)^3}\,e^{ixP^+z^--i\vec k_T\cdot\vec z_T}\,\Gamma^{+R;R+}(z)&=2\,\mathcal O^g_{+1-1}\equiv\mathcal O^g_{T_R},\\
\int\frac{\ud z^-\,\ud^2z_T}{(2\pi)^3}\,e^{ixP^+z^--i\vec k_T\cdot\vec z_T}\,\Gamma^{+L;L+}(z)&=2\,\mathcal O^g_{-1+1}\equiv\mathcal O^g_{T_L},
\end{align}
where for the gluon polarization vectors we used
\begin{eqnarray}
\vec \epsilon_{+1}=-\frac{1}{\sqrt{2}}\,(1,i,0),\qquad\vec \epsilon_{-1}=\frac{1}{\sqrt{2}}\,(1,-i,0).
\end{eqnarray}
Denoting the matrix elements of these leading-twist operators as follows
\begin{align}
\langle p',\Lambda'|\mathcal O^{q,g}_U|p,\Lambda\rangle&\equiv U^{q,g},\\
\langle p',\Lambda'|\mathcal O^{q,g}_L|p,\Lambda\rangle&\equiv L^{q,g},\\
\langle p',\Lambda'|\mathcal O^{q,g}_{T_R}|p,\Lambda\rangle&\equiv T^{q,g}_R,\\
\langle p',\Lambda'|\mathcal O^{q,g}_{T_L}|p,\Lambda\rangle&\equiv T^{q,g}_L,
\end{align}
we obtain the following matrix representation for the light-front helicity amplitudes
\begin{align}\label{Aq}
H^{q,g}_{\Lambda'\lambda',\Lambda\lambda}&=\left(
\begin{array}{c|c}
\frac{1}{2}(U^{q,g}+L^{q,g})&\tfrac{1}{2}\,T_R^{q,g}\\\hline
\tfrac{1}{2}\,T_L^{q,g}&\frac{1}{2}(U^{q,g}-L^{q,g})
\end{array}
\right),
\end{align}
where the row entries are $\lambda'=+J,-J$ and the column entries are likewise $\lambda=+J,-J$ with $J=\tfrac{1}{2}$ for quarks and $J=1$ for gluons. Each inner block in eq.~\eqref{Aq} is a $2\times 2$ matrix in the space of nucleon light-front helicity, as specified in eqs.~\eqref{Mat0}, \eqref{Mat1} and \eqref{Mat2}. 

Using the discrete symmetry and hermiticity constraints discussed in section~\ref{symmetries}, one obtains the following properties for the helicity amplitudes:
\begin{align}
\text{Hermiticity}\qquad H_{\Lambda'\lambda',\Lambda\lambda}(P,k,\Delta,N;\eta_i)&=H^*_{\Lambda\lambda,\Lambda'\lambda'}(P,k,-\Delta,N;\eta_i),\\
\text{LF Parity}\qquad H_{\Lambda'\lambda',\Lambda\lambda}(P,k,\Delta,N;\eta_i)&=H_{-\Lambda'-\lambda',-\Lambda-\lambda}(\tilde P,\tilde k,\tilde\Delta,\tilde N;\eta_i),\label{eq:parity-hel}\\
\text{LF Time-reversal}\qquad H_{\Lambda'\lambda',\Lambda\lambda}(P,k,\Delta,N;\eta_i)&=(-1)^{\Delta\ell_z}H^*_{\Lambda'\lambda',\Lambda\lambda}(\tilde P,\tilde k,\tilde\Delta,N;-\eta_i).\label{TR-hel}
\end{align}
Explicit calculation gives for the quark light-front helicity amplitudes at twist 2:
{\allowdisplaybreaks
\begin{align}\label{qhel1}
H^{q}_{+\tfrac{1}{2}+\tfrac{1}{2},+\tfrac{1}{2}+\tfrac{1}{2}}&=\frac{1}{2}\left[\left(S^{0,+;q}_{1,1a}+S^{0,-;q}_{1,1a}\right)+\frac{i(\vec k_T\times\vec\Delta_T)_z}{M^2}\left(S^{0,+;q}_{1,1b}+S^{0,-;q}_{1,1b}\right)\right],\\
H^{q}_{-\tfrac{1}{2}+\tfrac{1}{2},-\tfrac{1}{2}+\tfrac{1}{2}}&=\frac{1}{2}\left[\left(S^{0,+;q}_{1,1a}-S^{0,-;q}_{1,1a}\right)-\frac{i(\vec k_T\times\vec\Delta_T)_z}{M^2}\left(S^{0,+;q}_{1,1b}-S^{0,-;q}_{1,1b}\right)\right],\\
H^{q}_{+\tfrac{1}{2}+\tfrac{1}{2},-\tfrac{1}{2}+\tfrac{1}{2}}&=\frac{1}{2}
\left[-\frac{k_L}{M}\left(P^{0,+;q}_{1,1a}-P^{0,-;q}_{1,1a}\right)-\frac{ \Delta_L}{M}\left(P^{0,+;q}_{1,1b}-P^{0,-;q}_{1,1b}\right)\right],\\
H^{q}_{-\tfrac{1}{2}+\tfrac{1}{2},+\tfrac{1}{2}+\tfrac{1}{2}}&=\frac{1}{2}
\left[\frac{k_R}{M}\left(P^{0,+;q}_{1,1a}+P^{0,-;q}_{1,1a}\right)+\frac{ \Delta_R}{M}\left(P^{0,+;q}_{1,1b}+P^{0,-;q}_{1,1b}\right)\right],\\
H^{q}_{+\tfrac{1}{2}+\tfrac{1}{2},+\tfrac{1}{2}-\tfrac{1}{2}}&=\frac{1}{2}
\left[\frac{k_R}{M}\left(P^{1,-;q}_{1,1a}+P'^{1,-;q}_{1,1a}\right)+\frac{\Delta_R}{M}\left(P^{1,-;q}_{1,1b}+P'^{1,-;q}_{1,1b}\right)\right],\\
H^{q}_{-\tfrac{1}{2}+\tfrac{1}{2},-\tfrac{1}{2}-\tfrac{1}{2}}&=\frac{1}{2}
\left[\frac{k_R}{M}\left(P^{1,-;q}_{1,1a}-P'^{1,-;q}_{1,1a}\right)+\frac{\Delta_R}{M}\left(P^{1,-;q}_{1,1b}-P'^{1,-;q}_{1,1b}\right)\right],\\
H^{q}_{+\tfrac{1}{2}+\tfrac{1}{2},-\tfrac{1}{2}-\tfrac{1}{2}}&=\frac{1}{2}
\left[S^{1,-;q}_{1,1a}+\frac{i(\vec k_T\times\vec\Delta_T)_z}{M^2}\, S^{1,-;q}_{1,1b}\right],\\
H^{q}_{-\tfrac{1}{2}+\tfrac{1}{2},+\tfrac{1}{2}-\tfrac{1}{2}}&=\frac{1}{2}
\left[\frac{k_R^2}{M^2}\,D^{1,-;q}_{1,1a}+\frac{\Delta_R^2}{M^2}\,D^{1,-;q}_{1,1b}\right].
\label{qhel8}
\end{align}
}
Similarly, for the gluon helicity amplitudes at twist 2, we have
{\allowdisplaybreaks
\begin{align}\label{ghel1}
H^{g}_{+\tfrac{1}{2}+1,+\tfrac{1}{2}+1}&=\frac{1}{2}\left[\left(S^{0,+;g}_{1,1a}+S^{0,-;g}_{1,1a}\right)+\frac{i(\vec k_T\times\vec\Delta_T)_z}{M^2}\left(S^{0,+;g}_{1,1b}+S^{0,-;g}_{1,1b}\right)\right],\\
\label{ghel2}
H^{g}_{-\tfrac{1}{2}+1,-\tfrac{1}{2}+1}&=\frac{1}{2}\left[\left(S^{0,+;g}_{1,1a}-S^{0,-;g}_{1,1a}\right)-\frac{i(\vec k_T\times\vec\Delta_T)_z}{M^2}\left(S^{0,+;g}_{1,1b}-S^{0,-;g}_{1,1b}\right)\right],\\
\label{ghel3}
H^{g}_{+\tfrac{1}{2}+1,-\tfrac{1}{2}+1}&=\frac{1}{2}
\left[-\frac{k_L}{M}\left(P^{0,+;g}_{1,1a}-P^{0,-;g}_{1,1a}\right)-\frac{ \Delta_L}{M}\left(P^{0,+;g}_{1,1b}-P^{0,-;g}_{1,1b}\right)\right],\\
\label{ghel4}
H^{g}_{-\tfrac{1}{2}+1,\tfrac{1}{2}+1}&=\frac{1}{2}
\left[\frac{k_R}{M}\left(P^{0,+;g}_{1,1a}+P^{0,-;g}_{1,1a}\right)+\frac{ \Delta_R}{M}\left(P^{0,+;g}_{1,1b}+P^{0,-;g}_{1,1b}\right)\right],\\
\label{ghel5}
H^{g}_{+\tfrac{1}{2}+1,+\tfrac{1}{2}-1}&=-\frac{1}{2}
\left[\frac{k^{2}_{R}}{M^2}\left(D^{2,+;g}_{1,1a}+D'^{2,+;g}_{1,1a}\right)+\frac{\Delta^{2}_{R}}{M^2}\left(D^{2,+;g}_{1,1b}+D'^{2,+;g}_{1,1b}\right)\right],\\
\label{ghel6}
H^{g}_{-\tfrac{1}{2}+1,-\tfrac{1}{2}-1}&=-\frac{1}{2}
\left[\frac{k^{2}_{R}}{M^2}\left(D^{2,+;g}_{1,1a}-D'^{2,+;g}_{1,1a}\right)+\frac{\Delta^{2}_{R}}{M^2}\left(D^{2,+;g}_{1,1b}-D'^{2,+;g}_{1,1b}\right)\right],\\
\label{ghel7}
H^{g}_{+\tfrac{1}{2}+1,-\tfrac{1}{2}-1}&=-\frac{1}{2}
\left[\frac{k_R}{M}\,P^{2,+;g}_{1,1a}+\frac{\Delta_R}{M}\,P^{2,+;g}_{1,1b}\right],\\
\label{ghel8}
H^{g}_{-\tfrac{1}{2}+1,+\tfrac{1}{2}-1}&=-\frac{1}{2}
\left[\frac{k^{3}_{R}}{M^3}\,F^{2,+;g}_{1,1a}+\frac{\Delta^{3}_{R}}{M^3}\,F^{2,+;g}_{1,1b}\right].
\end{align}
}

\section{Projections of GTMDs onto TMDs and GPDs}
\label{section-4}
\subsection{TMD limit}

The forward limit $\Delta=0$ of the correlators $W$, denoted as $\Phi$,
\begin{align}
\Phi^{[\Gamma]}_{\Lambda'\Lambda}(P,x,\vec k_T,N;\eta)&=W^{[\Gamma]}_{\Lambda'\Lambda}(P,x,\vec k_T,0,N;\eta)\nonumber\\
&\hspace{-3cm}=\frac{1}{2}\int\frac{\ud z^-\,\ud^2z_T}{(2\pi)^3}\,e^{ixP^+z^--i\vec k_T\cdot\vec z_T}\,\langle P,\Lambda'|\overline\psi(-\tfrac{z}{2})\Gamma\,\mathcal W\,\psi(\tfrac{z}{2})|P,\Lambda\rangle\Big|_{z^+=0},\\
\Phi^{\mu\nu;\rho\sigma}_{\Lambda'\Lambda}(P,x,\vec k_T,N;\eta,\eta')&=W^{\mu\nu;\rho\sigma}_{\Lambda'\Lambda}(P,x,\vec k_T,0,N;\eta,\eta')\nonumber\\
&\hspace{-3cm}=\frac{1}{xP^+}\int\frac{\ud z^-\,\ud^2z_T}{(2\pi)^3}\,e^{ixP^+z^--i\vec k_T\cdot\vec z_T}\,\langle P,\Lambda'|2\uTr\!\left[G^{\mu\nu}(-\tfrac{z}{2})\,\mathcal W\,G^{\rho\sigma}(\tfrac{z}{2})\,\mathcal W'\right]|P,\Lambda\rangle\Big|_{z^+=0},
\label{tmd-gluon}
\end{align}
gives the quark-quark and gluon-gluon correlators which are parametrized in terms of quark and gluon TMDs, respectively. These TMDs can be seen as the forward limit of the GTMDs. For $\Delta =0$, the imaginary part of the GTMDs belonging to the class $X_+$ and the real part of the GTMDs belonging to the class $X_-$ vanish because they are odd under a sign change of $\Delta$, see eqs. \eqref{x1} and \eqref{x2}. In addition, the functions in eqs.~\eqref{P1}-\eqref{P5} which are multiplied by a coefficient proportional to $\Delta$, \emph{i.e.} those labeled by $b$, do not appear in the correlator $\Phi$ any longer.

In the quark sector, we find that in the TMD limit up to twist 4 only 32 distributions survive, in agreement with the results of refs.~\cite{Goeke:2005hb,Meissner:2009ww}. We provide here the relations of these TMDs with the GTMDs:
\begin{align}
\label{tmd1}
f^q_1&=\Re\textrm{e}\!\left[S^{0,+;q}_{1,1a}(x,0,\vec k^2_T,0,0;\eta)\right],& f^{\perp q}_{1T}&=-\Im\textrm{m}\!\left[P^{0,+;q}_{1,1a}(x,0,\vec k^2_T,0,0;\eta)\right],\\
\label{tmd2}
g^q_{1L}&=\Re\textrm{e}\!\left[S^{0,-;q}_{1,1a}(x,0,\vec k^2_T,0,0;\eta)\right],& g^q_{1T}&=\Re\textrm{e}\!\left[P^{0,-;q}_{1,1a}(x,0,\vec k^2_T,0,0;\eta)\right],\\
\label{tmd3}
h^q_1&=\tfrac{1}{2}\,\Re\textrm{e}\!\left[S^{1,-;q}_{1,1a}(x,0,\vec k^2_T,0,0;\eta)\right],& h^{\perp q}_1&=-\Im\textrm{m}\!\left[P^{1,-;q}_{1,1a}(x,0,\vec k^2_T,0,0;\eta)\right],\\
\label{tmd4}
h^{\perp q}_{1L}&=\Re\textrm{e}\!\left[P'^{1,-;q}_{1,1a}(x,0,\vec k^2_T,0,0;\eta)\right],& h^{\perp q}_{1T}&=\Re\textrm{e}\!\left[D^{1,-;q}_{1,1a}(x,0,\vec k^2_T,0,0;\eta)\right],\\
\label{tmd5}
e^q&=\Re\textrm{e}\!\left[S^{0,+;q}_{2,1a}(x,0,\vec k^2_T,0,0;\eta)\right],& e^{\perp q}_T&=-\Im\textrm{m}\!\left[P^{0,+;q}_{2,1a}(x,0,\vec k^2_T,0,0;\eta)\right],\\
\label{tmd6}
e^q_L&=-\Im\textrm{m}\!\left[S^{0,-;q}_{2,2a}(x,0,\vec k^2_T,0,0;\eta)\right],& e^q_T&=-\Im\textrm{m}\!\left[P^{0,-;q}_{2,2a}(x,0,\vec k^2_T,0,0;\eta)\right],\\
\label{tmd7}
f^q_T&=-\tfrac{1}{2}\,\Im\textrm{m}\!\left[S^{1,+;q}_{2,1a}(x,0,\vec k^2_T,0,0;\eta)\right],& f^{\perp q}&=\Re\textrm{e}\!\left[P^{1,+;q}_{2,1a}(x,0,\vec k^2_T,0,0;\eta)\right], \\
\label{tmd8}
f^{\perp q}_{L}&=-\Im\textrm{m}\!\left[P'^{1,+;q}_{2,1a}(x,0,\vec k^2_T,0,0;\eta)\right],& f^{\perp q}_T&=\Im\textrm{m}\!\left[D^{1,+;q}_{2,1a}(x,0,\vec k^2_T,0,0;\eta)\right], \\
g^q_T&=\tfrac{1}{2}\,\Re\textrm{e}\!\left[S^{1,-;q}_{2,1a}(x,0,\vec k^2_T,0,0;\eta)\right],& g^{\perp q}&=-\Im\textrm{m}\!\left[P^{1,-;q}_{2,1a}(x,0,\vec k^2_T,0,0;\eta)\right],\\
g^{\perp q}_L&=\Re\textrm{e}\!\left[P'^{1,-;q}_{2,1a}(x,0,\vec k^2_T,0,0;\eta)\right],& g^{\perp q}_T&=\Re\textrm{e}\!\left[D^{1,-;q}_{2,1a}(x,0,\vec k^2_T,0,0;\eta)\right],\\
h^q&=-\Im\textrm{m}\!\left[S^{0,+;q}_{2,2a}(x,0,\vec k^2_T,0,0;\eta)\right],& 
h^{\perp q}_T&=\Re\textrm{e}\!\left[P^{0,+;q}_{2,2a}(x,0,\vec k^2_T,0,0;\eta)\right],\\
h^q_L&=\Re\textrm{e}\!\left[S^{0,-;q}_{2,1a}(x,0,\vec k^2_T,0,0;\eta)\right],& 
h^q_T&=\Re\textrm{e}\!\left[P^{0,-;q}_{2,1a}(x,0,\vec k^2_T,0,0;\eta)\right],\\
f^q_3&=\Re\textrm{e}\!\left[S^{0,+;q}_{3,1a}(x,0,\vec k^2_T,0,0;\eta)\right],& 
f^{\perp q}_{3T}&=-\Im\textrm{m}\!\left[P^{0,+;q}_{3,1a}(x,0,\vec k^2_T,0,0;\eta)\right],\\
g^q_{3L}&=\Re\textrm{e}\!\left[S^{0,-;q}_{3,1a}(x,0,\vec k^2_T,0,0;\eta)\right],& g^q_{3T}&=\Re\textrm{e}\!\left[P^{0,-;q}_{3,1a}(x,0,\vec k^2_T,0,0;\eta)\right],\\
h^q_3&=\tfrac{1}{2}\,\Re\textrm{e}\!\left[S^{1,-;q}_{3,1a}(x,0,\vec k^2_T,0,0;\eta)\right],& h^{\perp q}_3&=-\Im\textrm{m}\!\left[P^{1,-;q}_{3,1a}(x,0,\vec k^2_T,0,0;\eta)\right],\\
h^{\perp q}_{3L}&=\Re\textrm{e}\!\left[P'^{1,-;q}_{3,1a}(x,0,\vec k^2_T,0,0;\eta)\right],& h^{\perp q}_{3T}&=\Re\textrm{e}\!\left[D^{1,-;q}_{3,1a}(x,0,\vec k^2_T,0,0;\eta)\right].
\end{align}
The 12 TMDs given by the imaginary part of the GTMDs are T-odd, while the other 20 given by the real part of the GTMDs are T-even. Using the definitions \cite{Avakian:2010br}
\begin{align}
h^q_1&=h^q_{1T}+\frac{\vec k^2_T}{2M^2}\,h^{\perp q}_{1T},\\
f^q_T&=f'^q_T+\frac{\vec k^2_T}{2M^2}\,f^{\perp q}_T,\\
g^q_T&=g'^q_T+\frac{\vec k^2_T}{2M^2}\,g^{\perp q}_T,\\
h^q_3&=h^q_{3T}+\frac{\vec k^2_T}{2M^2}\,h^{\perp q}_{3T},
\end{align}
together with the results in appendix~\ref{app:1} relating the quark GTMDs introduced in this work to the ones of ref.~\cite{Meissner:2009ww}, we reproduce the TMD limit of the quark GTMDs given by the eqs. (4.3)-(4.34) of ref.~\cite{Meissner:2009ww}.

In the gluon sector, we find 8 TMDs at twist 2, 16 TMDs at twist 3, 24 TMDs at twist 4. The correlators at twist 5 and twist 6 are copies of the correlators at twist 3 and 2, respectively. At each twist, half of the TMDs are T-odd functions and half are T-even functions. We will discuss explicitly the parametrizations for the gluon correlators at twist 2 and at twist 3, comparing with the results derived in refs.~\cite{Mulders:2000sh,Goeke:2005hb}.
 
We introduce the covariant light-front spin vector $S^\mu=[S_\parallel\tfrac{P^+}{M}, -S_\parallel\tfrac{P^-}{M},\vec S_T]$, which leads to the linear combination~\cite{Meissner:2007rx} 
\begin{equation}
\begin{aligned}
\Phi^{\mu\nu;\rho\sigma}
(P,x,\vec k_T,N;\eta|S)&=
\frac{1+S_\parallel}{2}\,\Phi^{\mu\nu;\rho\sigma}_{++}(P,x,\vec k_T,N;\eta)+\frac{1-S_\parallel}{2}\,\Phi^{\mu\nu;\rho\sigma}_{--}(P,x,\vec k_T,N;\eta)\\
&\phantom{=}+\frac{S_L}{2}\,\Phi^{\mu\nu;\rho\sigma}_{-+}(P,x,\vec k_T,N;\eta)+\frac{S_R}{2}\,\Phi^{\mu\nu;\rho\sigma}_{+-}(P,x,\vec k_T,N;\eta).
\end{aligned}
\end{equation}
Using the conventions of ref.~\cite{Meissner:2007rx}, the twist-2 gluon TMDs parametrize the gluon correlators as
\begin{align}
\label{eq1:gt-t2}
\delta_T^{ij}\Phi^{+i;+j}(P,x,\vec k_T,N;\eta|S) &=f^g_1(x,\vec k^2_T)-\frac{(\vec k_T\times\vec S_T)_z}{M}\,f^{\perp g}_{1T}(x,\vec k^2_T),\\
-i\epsilon_{T}^{ij}\Phi^{+i;+j}(P,x,\vec k_T,N;\eta|S)&=S_\parallel\,g^g_{1L}(x,\vec k^2_T)+\frac{\vec k_T\cdot\vec S_T}{M}\,g^g_{1T}(x,\vec k^2_T),\\
\Phi^{+R;+R}(P,x,\vec k_T,N;\eta|S)&=\frac{k^2_R}{2M^2}\,h^{\perp g}_1(x,\vec k^2_T)-\frac{k_R\,\epsilon^{Rk_T}_T}{2M^2}\,S_\parallel \,h^{\perp g}_{1L}(x,\vec k^2_T)\nonumber\\
&\hspace{-0.5cm}-\frac{k_R\,\epsilon^{RS_T}_T}{2M}\,h^g_{1T}(x,\vec k^2_T)-\frac{k_R\,\epsilon^{Rk_T}_T\,(\vec k_T\cdot\vec S_T)}{2M^3}\,h^{\perp g}_{1T}(x,\vec k^2_T).\label{eq3:gt-t2}
\end{align}
The relations between the leading-twist gluon TMDs and GTMDs read
\begin{align}
\label{gtmd1}
f_1^g&=\Re\textrm{e}\!\left[S^{0,+;g}_{1,1a}(x,0,\vec k^2_T,0,0;\eta)\right],
& f^{\perp g}_{1T}&=-\Im\textrm{m}\!\left[P^{0,+;g}_{1,1a}(x,0,\vec k^2_T,0,0;\eta)\right],\\
\label{gtmd2}
g^g_{1L}&=\Re\textrm{e}\!\left[S^{0,-;g}_{1,1a}(x,0,\vec k^2_T,0,0;\eta)\right],
& g^g_{1T}&=\Re\textrm{e}\!\left[P^{0,-;g}_{1,1a}(x,0,\vec k^2_T,0,0;\eta)\right],\\
\label{gtmd3}
h^g_1&=\Im\textrm{m}\!\left[P^{2,+;g}_{1,1a}(x,0,\vec k^2_T,0,0;\eta)\right],
& h^{\perp g}_1&=2\,\Re\textrm{e}\!\left[D^{2,+;g}_{1,1a}(x,0,\vec k^2_T,0,0;\eta)\right],\\
\label{gtmd4}
h_{1L}^{\perp g}&=2\,\Im\textrm{m}\!\left[D'^{2,+;g}_{1,1a}(x,0,\vec k^2_T,0,0;\eta)\right],
&h^{\perp g}_{1T}&=2\,\Im\textrm{m}\!\left[F^{2,+;g}_{1,1a}(x,0,\vec k^2_T,0,0;\eta)\right],
\end{align}
where $h^g_1=h^g_{1T}+\frac{\vec k^2_T}{2M^2}\,h^{\perp g}_{1T}$. The twist-2 gluon TMDs  are related to those introduced in ref.~\cite{Mulders:2000sh} through 
 \begin{align}
f^g_1&=G,&f^{\perp g}_{1T}&=-G_T,\\
g^g_{1L}&=-\Delta G_L,&g^g_{1T}&=-\Delta G_T,\\
h^g_1&=-\Delta H_T,&h^{\perp g}_1&=H^\perp,\\
h_{1L}^{\perp g}&=-\Delta H^\perp_L,&h^{\perp g}_{1T}&=-\Delta H^\perp_T.
 \end{align}

For the  gluon correlators at twist 3, we follow the conventions of the corresponding quark correlators~\cite{Avakian:2010br} at the same twist order and  with the same values of $\Delta S_z$ and $c_{\mathsf P}$ (see tables~\ref{Qstruct} and \ref{Gstruct}). As a result we have
\begin{align}
\Phi^{\{+-;+R\}}(P,x,\vec k_T,N;\eta|S)&=\frac{M}{P^+}\left[\frac{k_R}{M}\,f^{\perp g}(x,\vec k^2_T)+\frac{\epsilon^{Rk_T}_T}{M}\,S_\parallel\,f^{\perp g}_L(x,\vec k^2_T)\right.\nonumber\\
&\hspace{-0.5cm}+\epsilon^{RS_T}_T f^g_T(x,\vec k^2_T)\left.+\frac{(k_Rk^j_T-\tfrac{1}{2}\,\vec k^2_T\,\delta^{Rj}_T)\,\epsilon^{jS_T}_T}{M^2}\,f^{\perp  g}_T(x,\vec k^2_T)\right],\\
\Phi^{[+-;+R]}(P,x,\vec k_T,N;\eta|S)&=\frac{iM}{P^+}\left[\frac{k_R}{M}\,\bar f^{\perp g}(x,\vec k^2_T)+\frac{\epsilon^{Rk_T}_T}{M}\,S_\parallel\,\bar f^{\perp g}_L(x,\vec k^2_T)\right.\nonumber\\
&\hspace{-0.5cm}+\epsilon^{RS_T}_T \bar f^g_T(x,\vec k^2_T)\left.+\frac{(k_Rk^j_T-\tfrac{1}{2}\,\vec k^2_T\,\delta^{Rj}_T)\,\epsilon^{jS_T}_T}{M^2}\,\bar f^{\perp  g}_T(x,\vec k^2_T)\right],\\
\tfrac{1}{2}\,\Phi^{[LR;+R]}(P,x,\vec k_T,N;\eta|S)&=\frac{M}{P^+}\left[\frac{\epsilon^{Rk_T}_T}{M}\,g^{\perp g}(x,\vec k^2_T)+\frac{k_R}{M}\,S_\parallel\,g_L^{\perp g}(x,\vec k^2_T)\right.\nonumber\\
&\hspace{-0.5cm}\left.+S_R \,g^g_T(x,\vec k^2_T)+\frac{(k_Rk^j_T-\tfrac{1}{2}\,\vec k^2_T\,\delta^{Rj}_T)\,S^j_T}{M^2}\,g_T^{\perp  g} (x,\vec k^2_T)\right],\\
\tfrac{1}{2}\,\Phi^{\{LR;+R\}}(P,x,\vec k_T,N;\eta|S)&=\frac{iM}{P^+}\left[\frac{\epsilon^{Rk_T}_T}{M}\,\bar g^{\perp g}(x,\vec k^2_T)+\frac{k_R}{M}\,S_\parallel\,\bar g_L^{\perp g}(x,\vec k^2_T)\right.\nonumber\\
&\hspace{-0.5cm}\left.+S_R \,\bar g^g_T(x,\vec k^2_T)+\frac{(k_Rk^j_T-\tfrac{1}{2}\,\vec k^2_T\,\delta^{Rj}_T)\,S^j_T}{M^2}\,\bar g_T^{\perp  g} (x,\vec k^2_T)\right].
\end{align}
The relations between the twist-3 gluon TMDs and GTMDs read 
\begin{align}\label{gtmd7}
f^g_T&=-\tfrac{1}{2}\,\Im\textrm{m}\!\left[S^{1,+;g}_{2,1a}(x,0,\vec k^2_T,0,0;\eta)\right],
&f^{\perp g}&=\Re\textrm{e}\!\left[P^{1,+;g}_{2,1a}(x,0,\vec k^2_T,0,0;\eta)\right],\\
f^{\perp g}_L&=-\Im\textrm{m}\!\left[P'^{1,+;g}_{2,1a}(x,0,\vec k^2_T,0,0;\eta)\right],
&f^{\perp g}_T&=\Im\textrm{m}\!\left[D^{1,+;g}_{2,1a}(x,0,\vec k^2_T,0,0;\eta)\right],\\
\bar f^g_T&=\tfrac{1}{2}\,\Re\textrm{e}\!\left[S^{1,+;g}_{2,2a}(x,0,\vec k^2_T,0,0;\eta)\right],
&\bar f^{\perp g}&=\Im\textrm{m}\!\left[P^{1,+;g}_{2,2a}(x,0,\vec k^2_T,0,0;\eta)\right],\\
\bar f^{\perp g}_L&=\Re\textrm{e}\!\left[P'^{1,+;g}_{2,2a}(x,0,\vec k^2_T,0,0;\eta)\right],
&\bar f^{\perp g}_T&=-\Re\textrm{e}\!\left[D^{1,+;g}_{2,2a}(x,0,\vec k^2_T,0,0;\eta)\right],\\
g^g_T&=\tfrac{1}{2}\,\Re\textrm{e}\!\left[S^{1,-;g}_{2,1a}(x,0,\vec k^2_T,0,0;\eta)\right],
&g^{\perp g}&=-\Im\textrm{m}\!\left[P^{1,-;g}_{2,1a}(x,0,\vec k^2_T,0,0;\eta)\right],\\
g^{\perp g}_L&=\Re\textrm{e}\!\left[P'^{1,-;g}_{2,1a}(x,0,\vec k^2_T,0,0;\eta)\right],
&g^{\perp g}_T&=\Re\textrm{e}\!\left[D^{1,-;g}_{2,1a}(x,0,\vec k^2_T,0,0;\eta)\right],\\
\bar g^g_T&=\tfrac{1}{2}\,\Im\textrm{m}\!\left[S^{1,-;g}_{2,2a}(x,0,\vec k^2_T,0,0;\eta)\right],
&\bar g^{\perp g}&=\Re\textrm{e}\!\left[P^{1,-;g}_{2,2a}(x,0,\vec k^2_T,0,0;\eta)\right],\\
\bar g^{\perp g}_L&=\Im\textrm{m}\!\left[P'^{1,-;g}_{2,2a}(x,0,\vec k^2_T,0,0;\eta)\right],
&\bar g^{\perp g}_T&=\Im\textrm{m}\!\left[D^{1,-;g}_{2,2a}(x,0,\vec k^2_T,0,0;\eta)\right].
\label{tmd-last}
\end{align}
The twist-3 gluon TMDs  are related to those introduced in ref.~\cite{Mulders:2000sh} through 
\begin{align}\label{gtmd7}
f^g_T&=-\tfrac{1}{2}\,\Im\textrm{m}\!\left[\Delta G_{3T}\right],
&f^{\perp g}&=\tfrac{1}{2}\,\Re\textrm{e}\!\left[G^\perp_3\right],\\
f^{\perp g}_L&=-\tfrac{1}{2}\,\Im\textrm{m}\!\left[\Delta G^\perp_{3L}\right],
&f^{\perp g}_T&=\tfrac{1}{2}\,\Im\textrm{m}\!\left[\Delta G^\perp_{3T}\right],\\
\bar f^g_T&=-\tfrac{1}{2}\,\Re\textrm{e}\!\left[\Delta G_{3T}\right],
&\bar f^{\perp g}&=-\tfrac{1}{2}\,\Im\textrm{m}\!\left[G^\perp_3\right],\\
\bar f^{\perp g}_L&=-\tfrac{1}{2}\,\Re\textrm{e}\!\left[\Delta G^\perp_{3L}\right],
&\bar f^{\perp g}_T&=\tfrac{1}{2}\,\Re\textrm{e}\!\left[\Delta G^\perp_{3T}\right],\\
g^g_T&=\tfrac{1}{2}\,\Re\textrm{e}\!\left[\Delta H_{3T}\right],
&g^{\perp g}&=\tfrac{1}{2}\,\Im\textrm{m}\!\left[H^\perp_3\right],\\
g^{\perp g}_L&=\tfrac{1}{2}\,\Re\textrm{e}\!\left[\Delta H^\perp_{3L}\right],
&g^{\perp g}_T&=\tfrac{1}{2}\,\Re\textrm{e}\!\left[\Delta H^\perp_{3T}\right],\\
\bar g^g_T&=\tfrac{1}{2}\,\Im\textrm{m}\!\left[\Delta H_{3T}\right],
&\bar g^{\perp g}&=-\tfrac{1}{2}\,\Re\textrm{e}\!\left[H^\perp_3\right],\\
\bar g^{\perp g}_L&=\tfrac{1}{2}\,\Im\textrm{m}\!\left[\Delta H^\perp_{3L}\right],
&\bar g^{\perp g}_T&=\tfrac{1}{2}\,\Im\textrm{m}\!\left[\Delta H^\perp_{3T}\right].
\label{tmd-last}
\end{align}

\subsection{GPD limit}
  
Integrating the correlator $W$ over $\vec k_T$, one obtains the parton correlators denoted as $F$
\begin{align}
F^{[\Gamma]}_{\Lambda'\Lambda}(P,x,\Delta,N)&=\int\ud^2k_T\,W^{[\Gamma]}_{\Lambda'\Lambda}(P,x,\vec k_T,\Delta,N;\eta)\nonumber\\
&=\frac{1}{2}\int\frac{\ud z^-}{2\pi}\,e^{ixP^+z^-}\,\langle p',\Lambda'|\overline\psi(-\tfrac{z^-}{2})\Gamma\,\mathcal W\,\psi(\tfrac{z^-}{2})|p,\Lambda\rangle,
\label{gpd-quark}\\
F^{\mu\nu;\rho\sigma}_{\Lambda'\Lambda}(P,x,\Delta,N)&=\int\ud^2k_T\,W^{\mu\nu;\rho\sigma}_{\Lambda'\Lambda}(P,x,\vec k_T,\Delta,N;\eta,\eta')\nonumber\\
&=\frac{1}{xP^+}\int\frac{\ud z^-}{2\pi}\,e^{ixP^+z^-}\,
\langle p',\Lambda'|2\uTr\!\left[G^{\mu\nu}(-\tfrac{z^-}{2})\,\mathcal W\,G^{\rho\sigma}(\tfrac{z^-}{2})\,\mathcal W'\right]|p,\Lambda\rangle.
\label{gpd-gluon}
\end{align}
The integration over $\vec k_T$ removes the dependence on $\eta_i$, and we are left with a Wilson line connecting directly the points $-\tfrac{z^-}{2}$ and $\tfrac{z^-}{2}$ by a straight line. As a consequence, all the T-odd contributions given by the imaginary part of the GTMDs disappear, and the generic structures parametrizing the correlators \eqref{gpd-quark}-\eqref{gpd-gluon} can be obtained from eqs.~\eqref{struc1}-\eqref{struc4}  as
\begin{align}
\int \ud^2k_T\,S&=\int \ud^2k_T\,\Re\textrm{e}\,S_{t,ia}\equiv\mathcal S_{t,i}(x,\xi,\vec \Delta^2_T),\\
\int \ud^2k_T\,P_{R(L)}&=\frac{\Delta_{R(L)}}{M}\int\ud^2k_T \left(\frac{\vec k_T\cdot\vec \Delta_T}{\vec\Delta^2_T}\,\Re\textrm{e}\,P_{t,ia}+\Re\textrm{e}\,P_{t,ib}\right)\nonumber\\
&=\frac{\Delta_{R(L)}}{M}\,\mathcal P_{t,i}(x,\xi,\vec \Delta^2_T),\\
\int \ud^2k_T\,{D}_{R(L)}&=\frac{\Delta^2_{R(L)}}{M^2}\int\ud^2k_T \left[\frac{2(\vec k_T\cdot\vec\Delta_T)^2-\vec k^2_T\,\vec\Delta^2_T}{(\vec\Delta^2_T)^2}\,
\Re\textrm{e}\,D_{t,ia}+\Re\textrm{e}\,D_{t,ib}\right]\nonumber\\
&=\frac{\Delta^2_{R(L)}}{M^2}\,\mathcal D_{t,i}(x,\xi,\vec\Delta^2_T),\\
\int \ud^2k_T\,{F}_{R(L)}&=\frac{\Delta^3_{R(L)}}{M^3}\int\ud^2k_T \left[\frac{\left(4(\vec k_T\cdot\vec\Delta_T)^2-3\,\vec k^2_T\,\vec\Delta^2_T\right)(\vec k_T\cdot\vec\Delta_T)}{(\vec\Delta^2_T)^3}\,
\Re\textrm{e}\,F_{t,ia}+\Re\textrm{e}\,F_{t,ib}\right]\nonumber\\
&=\frac{\Delta^3_{R(L)}}{M^3}\,\mathcal F_{t,i}(x,\xi,\vec\Delta^2_T).
\end{align}

We refer to~\cite{Meissner:2009ww} for the complete list of quark GPDs up to twist 4, where the results at twist 3 in the chiral-odd sector and at twist 4 have been derived for the first time, the results at twist 2 follow the common definitions~\cite{Diehl:2003ny}, and the definitions at twist 3 in the chiral-even sector can easily be related to the set of GPDs introduced in ref.~\cite{Kiptily:2002nx}. The relations between the standard GPDs and the GPD limit of our GTMDs read:
\begin{itemize}
\item[-] at twist 2, in the chiral-even sector
\begin{align}\label{gpd:1}
H^q&=\frac{1}{\sqrt{1-\xi^2}}\left[\mathcal S^{0,+;q}_{1,1}+2\xi^2\,\mathcal P^{0,+;q}_{1,1}\right],&E^q&=2\sqrt{1-\xi^2}\,\mathcal P^{0,+;q}_{1,1},\\
\tilde H^q&=\frac{1}{\sqrt{1-\xi^2}}\left[\mathcal S^{0,-;q}_{1,1}+2\xi\,\mathcal P^{0,-;q}_{1,1}\right],&\tilde E^q&=\frac{2\sqrt{1-\xi^2}}{\xi}\,\mathcal P^{0,-;q}_{1,1};
\end{align}
\item[-] at twist 2, in the chiral-odd sector
\begin{align}
H^q_T&=\frac{1}{2\sqrt{1-\xi^2}}\left[\mathcal S^{1,-;q}_{1,1}-4\xi\,\mathcal P'^{1,-;q}_{1,1}+\frac{\vec\Delta_{T}^2}{M^2}\,\mathcal D^{1,-;q}_{1,1}\right],\\
E^q_T&=\frac{2}{\sqrt{1-\xi^2}}\left[\mathcal P^{1,-;q}_{1,1}+\xi\,\mathcal P'^{1,-;q}_{1,1}+2\,\mathcal D^{1,-;q}_{1,1}\right],\\
\tilde H^q_T&=-2\sqrt{1-\xi^2}\,\mathcal D^{1,-;q}_{1,1},\\
\tilde E^q_T&=\frac{2}{\sqrt{1-\xi^2}}\left[\xi\,\mathcal P^{1,-;q}_{1,1}+\mathcal P'^{1,-;q}_{1,1}+2\xi\,\mathcal D^{1,-;q}_{1,1}\right];
\label{gpd:4}   
\end{align}  
\item[-] at twist 3, in the chiral-even sector
\begin{align}
H^q_2&=\frac{1}{\sqrt{1-\xi^2}}\left[\mathcal S^{0,+;q}_{2,1}+2\xi^2\,\mathcal P^{0,+;q}_{2,1}\right],
&E^q_2&=2\sqrt{1-\xi^2}\,\mathcal P^{0,+;q}_{2,1},\\
\tilde H^q_2&=\frac{1}{\sqrt{1-\xi^2}}\left[\mathcal S^{0,-;q}_{2,2}+2\xi\,\mathcal P^{0,-;q}_{2,2}\right],
&\tilde E^q_2&=-2\sqrt{1-\xi^2}\,\mathcal P^{0,-;q}_{2,2},\\
H'^q_2&=\frac{1}{\sqrt{1-\xi^2}}\left[\mathcal S^{0,+;q}_{2,2}+2\xi^2\,\mathcal P^{0,+;q}_{2,2}\right],
&E'^q_2&=2\sqrt{1-\xi^2}\,\mathcal P^{0,+;q}_{2,2},\\
\tilde H'^q_2&=\frac{1}{\sqrt{1-\xi^2}}\left[\mathcal S^{0,-;q}_{2,1}+2\xi\,\mathcal P^{0,-;q}_{2,1}\right],
&\tilde E'^q_2&=-2\sqrt{1-\xi^2}\,\mathcal P^{0,-;q}_{2,1};
\end{align}   
\item[-] at twist 3, in the chiral-odd sector
\begin{align}
H^q_{2T}&=\frac{1}{2\sqrt{1-\xi^2}}\left[\mathcal S^{1,+;q}_{2,1}-4\xi\,\mathcal P'^{1,+;q}_{2,1}+\frac{\vec\Delta^2_T}{M^2}\,\mathcal D_{2,1}^{1,+;q}\right],\\
E^q_{2T}&=\frac{2}{\sqrt{1-\xi^2}}\left[\mathcal P^{1,+;q}_{2,1}+\xi\,\mathcal P'^{1,+;q}_{2,1}+2\,\mathcal D^{1,+;q}_{2,1}\right],\\
\tilde H^q_{2T}&=-2\sqrt{1-\xi^2}\,\mathcal D^{1,+;q}_{2,1},\\
 \tilde E^q_{2T}&=\frac{2}{\sqrt{1-\xi^2}}\left[\xi\,\mathcal P^{1,+;q}_{2,1}+\mathcal P'^{1,+;q}_{2,1}+2\xi\,\mathcal D^{1,+;q}_{2,1}\right],\\  
H'^q_{2T}&=\frac{1}{2\sqrt{1-\xi^2}}\left[\mathcal S^{1,-;q}_{2,1}-4\xi\,\mathcal P'^{1,-;q}_{2,1}+\frac{\vec\Delta^2_T}{M^2}\,\mathcal D^{1,-;q}_{2,1}\right],\\
E'^q_{2T}&=\frac{2}{\sqrt{1-\xi^2}}\left[\mathcal P^{1,-;q}_{2,1}+\xi\,\mathcal P'^{1,-;q}_{2,1}+2\,\mathcal D^{1,-;q}_{2,1}\right],\\
\tilde H'^q_{2T}&=-2\sqrt{1-\xi^2}\,\mathcal D^{1,-;q}_{2,1},\\
\tilde E'^q_{2T}&=\frac{2}{\sqrt{1-\xi^2}}\left[\xi\,\mathcal P^{1,-;q}_{2,1}+\mathcal P'^{1,-;q}_{2,1}+2\xi\,\mathcal D^{1,-;q}_{2,1}\right];
\label{gpd:4}   
\end{align}  
\item[-] at twist 4, in the chiral-even sector
\begin{align}
H^q_3&=\frac{1}{\sqrt{1-\xi^2}}\left[\mathcal S^{0,+;q}_{3,1}+2\xi^2\,\mathcal P^{0,+;q}_{3,1}\right],&E^q_3&=2\sqrt{1-\xi^2}\,\mathcal P^{0,+;q}_{3,1},\\
\tilde H^q_3&=\frac{1}{\sqrt{1-\xi^2}}\left[\mathcal S^{0,-;q}_{3,1}+2\xi\,\mathcal P^{0,-;q}_{3,1}\right],&\tilde E^q_3&=\frac{2\sqrt{1-\xi^2}}{\xi}\,\mathcal P^{0,-;q}_{3,1};
\end{align}
  \item[-] at twist 4, in the chiral-odd sector
\begin{align}
H^q_{3T}&=\frac{1}{2\sqrt{1-\xi^2}}\left[\mathcal S^{1,-;q}_{3,1}-4\xi\,\mathcal P'^{1,-;q}_{3,1}+\frac{\vec\Delta_{T}^2}{M^2}\,\mathcal D^{1,-;q}_{3,1}\right],\\
E^q_{3T}&=\frac{2}{\sqrt{1-\xi^2}}\left[\mathcal P^{1,-;q}_{3,1}+\xi\,\mathcal P'^{1,-;q}_{3,1}+2\,\mathcal D^{1,-;q}_{3,1}\right],\\
\tilde H^q_{3T}&=-2\sqrt{1-\xi^2}\,\mathcal D^{1,-;q}_{3,1},\\
\tilde E^q_{3T}&=\frac{2}{\sqrt{1-\xi^2}}\left[\xi\,\mathcal P^{1,-;q}_{3,1}+\mathcal P'^{1,-;q}_{3,1}+2\xi\,\mathcal D^{1,-;q}_{3,1}\right]. 
\end{align}  
\end{itemize} 
Using the results in appendix~\ref{app:1} to relate the quark GTMDs introduced in this work and the ones of ref.~\cite{Meissner:2009ww}, we reproduce the GPD limit of the quark GTMDs given in eqs. (4.47)-(4.78) of ref.~\cite{Meissner:2009ww}. Using the hermiticity constraint~\eqref{xh} for the GTMDs, one  derives the symmetry behavior of the GPDs under the transformation $\xi\mapsto-\xi$. In the quark sector,  the 10 GPDs $\tilde E_T^q$, $\tilde H^q_{2}$, $H'^q_{2}$, $E'^q_{2}$, $\tilde E'^q_{2}$, $H^q_{2T}$,  $E^q_{2T}$, $\tilde{H}^q_{2T}$, $\tilde{E}'^q_{2T}$ and $\tilde{E}^{q}_{3T}$ are odd functions in $\xi$, while all the 22 other ones are even in $\xi$.
 
At twist 2, the gluon generalized correlators in the GPD limit are parametrized as~\cite{Meissner:2007rx} 
\begin{align} \label{e:ggpd1}
\delta_T^{ij}F^{+i;+j}_{\Lambda'\Lambda}(P,x,\Delta,N)&=\frac{1}{2P^+}\,\overline u(p',\Lambda')\left[\gamma^+\,H^g(x,\xi,t)+\frac{i\sigma^{+\mu}\Delta_\mu}{2M}\,E^g(x,\xi,t)\right]u(p,\Lambda),\\
\label{e:ggpd2}
-i\epsilon^{ij}_TF^{+i;+j}_{\Lambda'\Lambda}(P,x,\Delta,N)&=\frac{1}{2P^+}\,\overline u(p',\Lambda')\left[\gamma^+\gamma_5\,\tilde H^g(x,\xi,t)+\frac{\Delta^+\gamma_5}{2M}\,\tilde E^g(x,\xi,t)\right]u(p,\Lambda),\\
\label{e:ggpd3}
F^{+R;+R}_{\Lambda'\Lambda}(P,x,\Delta,N)&=\frac{1}{2P^+}\,\frac{\Delta^+P_R-P^+\Delta_R}{2MP^+}\nonumber\\
&\hspace{-1.5cm}\times\overline u(p',\Lambda')\left[i\sigma^{+R}\,H^g_T(x,\xi,t)+\frac{\gamma^+\Delta_R-\Delta^+\gamma_R}{2M}\,E^g_T(x,\xi,t)\right.\nonumber\\
&\hspace{-1.5cm}\left.+\frac{P^+\Delta_R-\Delta^+P_R}{M^2}\,\tilde H^g_T(x,\xi,t)+\frac{\gamma^+P_R-P^+\gamma_R}{M}\,\tilde E^g_T(x,\xi,t)\right]u(p,\Lambda).
\end{align}
The relations between these twist-2 gluon GPDs and the GPD limit of our GTMDs read:
\begin{itemize}
\item[-] in the chiral-even sector\begin{align}\label{g1}
H^g&=\frac{1}{\sqrt{1-\xi^2}}\left[\mathcal S^{0,+;g}_{1,1}+2\xi^2\,\mathcal P^{0,+;g}_{1,1}\right],
&E^g&=2\sqrt{1-\xi^2}\,\mathcal P^{0,+;g}_{1,1},\\
\tilde H^g&=\frac{1}{\sqrt{1-\xi^2}}\left[\mathcal S^{0,-;g}_{1,1}+2\xi\,\mathcal P^{0,-;g}_{1,1}\right],
&\tilde E^g&=\frac{2\sqrt{1-\xi^2}}{\xi}\,\mathcal P^{0,-;g}_{1,1};
\end{align}
\item[-] in the chiral-odd sector
\begin{align}
H^g_T&=-\frac{1}{\sqrt{1-\xi^2}}\left[\mathcal P^{2,+;g}_{1,1}-4\xi\,\mathcal D'^{2,+;g}_{1,1}+\frac{\vec\Delta^2_T}{M^2}\,\mathcal F^{2,+;g}_{1,1}\right],\\
E^g_T&=-\frac{4}{\sqrt{1-\xi^2}}\left[\mathcal D^{2,+;g}_{1,1}+\xi\,\mathcal D'^{2,+;g}_{1,1}+2\,\mathcal F^{2,+;q}_{1,1}\right],\\
\tilde H^g_T&=4\sqrt{1-\xi^2}\,\mathcal F^{2,+;g}_{1,1},\\
\tilde E^g_T&=-\frac{4}{\sqrt{1-\xi^2}}\left[\xi\,\mathcal D^{2,+;g}_{1,1}+\mathcal D'^{2,+;g}_{1,1}+2\xi\,\mathcal F^{2,+;g}_{1,1}\right].\label{gpd:g4} 
\end{align}  
\end{itemize}
The gluon GPDs at twist 3 are introduced here for the first time. For the  gluon correlators at twist 3, we follow the conventions of the corresponding quark correlators at the same twist order and with the same values of $\Delta S_z$ and $c_{\mathsf P}$ (see tables~\ref{Qstruct} and \ref{Gstruct}). Explicitly, the gluon GPDs at twist 3 can be defined according to
\begin{align}
F_{\Lambda'\Lambda}^{\{+-;+R\}}&=\frac{M}{2(P^+)^2}\,\overline u(p',\Lambda')\left[i\sigma^{+R}\,H^g_{2T}(x, \xi, t)+\frac{\gamma^+\Delta_R-\Delta^+\gamma_R}{2M}\,E^g_{2T}(x, \xi, t)\right. \nonumber\\
&\left.+\frac{P^+\Delta_R-\Delta^+P_R}{M^2}\,\tilde H^g_{2T}(x, \xi, t)+\frac{\gamma^+P_R-P^+\gamma_R}{M}\,\tilde E^g_{2T}(x, \xi, t)\right]u(p,\Lambda),\\
F_{\Lambda'\Lambda}^{[+-;+R]}&=\frac{M}{2(P^+)^2}\,\overline u(p',\Lambda')\left[i\sigma^{+R}\,\bar H^g_{2T}(x, \xi, t)+\frac{\gamma^+\Delta_R-\Delta^+\gamma_R}{2M}\,\bar E^g_{2T}(x, \xi, t)\right.\nonumber\\
&\left.+\frac{P^+\Delta_R-\Delta^+P_R}{M^2}\,\tilde{\bar{H}}^g_{2T}(x, \xi, t)+\frac{\gamma^+P_R-P^+\gamma_R}{M}\,\tilde{\bar{E}}^g_{2T}(x, \xi, t)\right]u(p,\Lambda),\\
\tfrac{1}{2}\,F_{\Lambda'\Lambda}^{[LR;+R]}&=\frac{M}{2(P^+)^2}\,\overline u(p',\Lambda')\left[i\sigma^{+R}\,H'^g_{2T}(x, \xi, t)+\frac{\gamma^+\Delta_R -\Delta^+\gamma_R}{2M}\,E'^g_{2T}(x, \xi, t)\right. \nonumber\\
&\left. +\frac{P^+\Delta_R-\Delta^+P_R}{M^2}\,\tilde H'^g_{2T}(x, \xi, t)+\frac{\gamma^+P_R-P^+\gamma_R}{M}\,\tilde E'^g_{2T}(x, \xi, t)\right]u(p,\Lambda),\\
\tfrac{1}{2}\,F_{\Lambda' \Lambda}^{\{LR;+R\}}&=\frac{M}{(P^+)^2}\,\overline u(p',\Lambda')\left[i\sigma^{+R}\,\bar H'^g_{2T}(x, \xi, t)+\frac{\gamma^+\Delta_R-\Delta^+\gamma_R}{2M}\,\bar E'^g_{2T}(x, \xi, t)\right. \nonumber\\
&\left.+\frac{P^+\Delta_R-\Delta^+P_R}{M^2}\,\tilde{\bar{H}}'^g_{2T}(x, \xi, t)+\frac{\gamma^+P_R-P^+\gamma_R}{M}\,\tilde{\bar{E}}'^g_{2T}(x, \xi, t)\right]u(p,\Lambda).
\end{align}
All the gluon GPDs at twist 3 are chiral-odd functions. The relations between these GPDs and the GPD limit of our GTMDs read:
\begin{align}
H^g_{2T}&=\frac{1}{2\sqrt{1-\xi^2}}\left[\mathcal S^{1,+;g}_{2,1}-4\xi\,\mathcal P'^{1,+;g}_{2,1}+\frac{\vec\Delta^2_T}{M^2}\,\mathcal D^{1,+;g}_{2,1}\right],\\
E^g_{2T}&=\frac{2}{\sqrt{1-\xi^2}}\left[\mathcal P^{1,+;g}_{2,1}+\xi\,\mathcal P'^{1,+;g}_{2,1}+2\,\mathcal D^{1,+;g}_{2,1}\right],\\
\tilde H^g_{2T}&=-2\sqrt{1-\xi^2}\,\mathcal D^{1,+;g}_{2,1},\\
\tilde E^g_{2T}&=\frac{2}{\sqrt{1-\xi^2}}\left[\xi\,\mathcal P^{1,+;g}_{2,1}+\mathcal P'^{1,+;g}_{2,1}+2\xi\,\mathcal D^{1,+;g}_{2,1}\right],\\  
\bar H^g_{2T}&=\frac{1}{2\sqrt{1-\xi^2}}\left[\mathcal S^{1,+;g}_{2,2}-4\xi\,\mathcal P'^{1,+;g}_{2,2}+\frac{\vec\Delta^2_T}{M^2}\,\mathcal D^{1,+;g}_{2,2}\right],\\
\bar E^g_{2T}&=\frac{2}{\sqrt{1-\xi^2}}\left[\mathcal P^{1,+;g}_{2,2}+\xi\,\mathcal P'^{1,+;g}_{2,2}+2\,\mathcal D^{1,+;g}_{2,2}\right],\\
\tilde{\bar{H}}^g_{2T}&=-2\sqrt{1-\xi^2}\,\mathcal D^{1,+;g}_{2,2},\\
\tilde{\bar{E}}^g_{2T}&=\frac{2}{\sqrt{1-\xi^2}}\left[\xi\,\mathcal P^{1,+;g}_{2,2}+\mathcal P'^{1,+;g}_{2,2}+2\xi\,\mathcal D^{1,+;g}_{2,2}\right],\\  
H'^g_{2T}&=\frac{1}{2\sqrt{1-\xi^2}}\left[\mathcal S^{1,-;g}_{2,1}-4\xi\,\mathcal P'^{1,-;g}_{2,1}+\frac{\vec\Delta^2_T}{M^2}\,\mathcal D^{1,-;g}_{2,1}\right],\\
E'^g_{2T}&=\frac{2}{\sqrt{1-\xi^2}}\left[\mathcal P^{1,-;g}_{2,1}+\xi\,\mathcal P'^{1,-;g}_{2,1}+2\,\mathcal D^{1,-;g}_{2,1}\right],\\
\tilde H'^g_{2T}&=-2\sqrt{1-\xi^2}\,\mathcal D^{1,-;g}_{2,1},\\
\tilde E'^g_{2T}&=\frac{2}{\sqrt{1-\xi^2}}\left[\xi\,\mathcal P^{1,-;g}_{2,1}+\mathcal P'^{1,-;g}_{2,1}+2\xi\,\mathcal D^{1,-;g}_{2,1}\right],\\  
\bar H'^g_{2T}&=\frac{1}{2\sqrt{1-\xi^2}}\left[\mathcal S^{1,-;g}_{2,2}-4\xi\,\mathcal P'^{1,-;g}_{2,2}+\frac{\vec\Delta^2_T}{M^2}\,\mathcal D^{1,-;g}_{2,2}\right],\\
\bar E'^g_{2T}&=\frac{2}{\sqrt{1-\xi^2}}\left[\mathcal P^{1,-;g}_{2,2}+\xi\,\mathcal P'^{1,-;g}_{2,2}+2\,\mathcal D^{1,-;g}_{2,2}\right],\\
\tilde{\bar{H}}'^g_{2T}&=-2\sqrt{1-\xi^2}\,\mathcal D^{1,-;g}_{2,2},\\
\tilde{\bar{E}}'^g_{2T}&=\frac{2}{\sqrt{1-\xi^2}}\left[\xi\,\mathcal P^{1,-;g}_{2,2}+\mathcal P'^{1,-;g}_{2,2}+2\xi\,\mathcal D^{1,-;g}_{2,2}\right].
\end{align}        
 From the hermiticity constraint~\eqref{xh}, one finds that the 9 gluon GPDs $\tilde E^g_T$, $H^g_{2T}$, $E^g_{2T}$, $\tilde H^g_{2T}$, $\tilde{\bar E}^g_{2T}$, $\tilde E'^g_{2T}$,  $\bar H'^g_{2T}$, $\bar E'^g_{2T}$ and $\tilde{\bar H}'^g_{2T}$ are odd functions in $\xi$, while all the 15 other ones are even in $\xi$.

\section{Conclusions}
\label{section-5}

We discussed the parametrization of the generalized off-diagonal two-parton correlators in terms of generalized transverse-momentum dependent parton distributions. Such distributions contain the most general information on the two-parton structure of hadrons and reduce  in specific limits or projections to the GPDs, TMDs and PDFs, and form factors accessible in various inclusive, semi-inclusive, exclusive, and elastic scattering processes. 

The structure of the generalized two-parton correlator has been analyzed by proposing a new method which can be applied in general to any matrix element of partonic operators and allows one to unravel the underlying spin and orbital angular momentum content. Such a method is based on the light-front formalism which provides the most natural and practical tools when dealing with distribution of partons in a fast moving hadron. We first give the classification of the parton operators in terms of $i)$ the spin-flip number, defined in terms of  the change of the light-front helicity and orbital angular momentum of the partons between the initial and final states, $ii)$  the properties under transformation by discrete symmetries, such as light-front parity and time-reversal, and $iii)$ the constraints from hermiticity. When calculating the off-diagonal matrix element of the parton operators between hadron states with given values of the light-front helicities and four-momentum, we can associate to each correlation function a unique multipole structure, related to the orbital angular momentum transferred to the hadrons. Such multipoles are then expressed in terms of powers of the average transverse momentum of the partons and the transverse momentum transferred to the hadrons, multiplied by Lorentz scalar functions representing the GTMDs. 

The method is applied simultaneously to the quark-quark and gluon-gluon correlation functions. In the quark sector, we obtain an alternative, but equivalent, parametrization to the one proposed in ref.~\cite{Meissner:2009ww} in terms of Lorentz covariant structures. The results for the gluon sector are presented here for the first time. We also discussed the GPD and TMD limit of the GTMDs, providing the relations with other existing parametrizations up to twist 3. The main advantage of the new nomenclature we propose  is to have a transparent and direct interpretation in terms of the spin and orbital angular momentum correlations encoded in each functions. This becomes particularly evident at leading twist, where the spin-flip number of the partonic operator can be identified with the difference of light-front helicities of the parton between the final and initial states, and therefore can be directly associated with a well-defined state of polarization of the parton. As outlined before, the proposed framework can be systematically  used for any matrix element of partonic operator and therefore provides a useful framework for the definition of new correlation functions that can be relevant for future phenomenological applications.
 
 \section*{Acknowledgments}

The authors acknowledge very kind and instructive discussions with A. Metz and P. Mulders. C. L. is also thankful to INFN and the Department of  Physics of the University of Pavia for the hospitality. This work was supported in part by  the European Community Joint Research Activity ``Study of Strongly Interacting Matter'' (acronym HadronPhysics3, Grant Agreement n. 283286) under the Seventh Framework Programme of the European Community, and by the P2I (``Physique des deux Infinis'') network.

\appendix

\section{Relations between different definitions of quark GTMDs }

\label{app:1}
In this appendix, we list the relations between  the quark GTMDs introduced in ref.~\cite{Meissner:2009ww} and the nomenclature adopted in this work.

We start with the parametrization of the quark correlator \eqref{gencorrq}  involving operators with $\Delta S_z=0$ and $c_{\mathsf P}=+1$ (third column of table~\ref{Qstruct}):
\begin{itemize}
\item[-] at twist 2, for $V^+$, we have
{\allowdisplaybreaks
\begin{align}
\label{rel1}
S^{0,+;q}_{1,1a}&=\frac{1}{\sqrt{1-\xi^2}}\,F_{1,1},\\
\label{rel2}
S^{0,+;q}_{1,1b}&=\frac{1}{\sqrt{1-\xi^2}}\,F_{1,4},\\
\label{rel3}
P^{0,+;q}_{1,1a}&=\sqrt{1-\xi^2}\,F_{1,2}+\frac{\xi}{\sqrt{1-\xi^2}}\,\frac{\vec \Delta_T^2}{2M^2}\,F_{1,4},\\
\label{rel4}
P^{0,+;q}_{1,1b}&=-\frac{1}{2\sqrt{1-\xi^2}}\,F_{1,1}+\sqrt{1-\xi^2}\,F_{1,3}-\frac{\xi}{\sqrt{1-\xi^2}}\,\frac{\vec k_T\cdot\vec \Delta_T}{2M^2}\,F_{1,4};
\end{align}
}
\item[-] at twist 3, for $S$, we have the same relations~\eqref{rel1}-\eqref{rel4} with the replacement $\{S^{0,+;q}_{1,1a},\, S^{0,+;q}_{1,1b},\, P^{0,+;q}_{1,1a},\, P^{0,+;q}_{1,1b}\}\mapsto\{S^{0,+;q}_{2,1a},\, S^{0,+;q}_{2,1b},\, P^{0,+;q}_{2,1a},\, P^{0,+;q}_{2,1b}\}$ on the left-hand side and  $\{F_{1,1},\, F_{1,2},\, F_{1,3},\, F_{1,4}\}\mapsto\{E_{2,1},\, E_{2,2},\, E_{2,3},\, E_{2,4}\}$ on the right-hand side;
\item[-] at twist 3, for $\tfrac{1}{2}\,T^{LR}$, we have the same relations~\eqref{rel1}-\eqref{rel4} with the substitution $\{S^{0,+;q}_{1,1a},\, S^{0,+;q}_{1,1b},\, P^{0,+;q}_{1,1a},\, P^{0,+;q}_{1,1b}\}\mapsto\{S^{0,+;q}_{2,2a},\, S^{0,+;q}_{2,2b},\, P^{0,+;q}_{2,2a},\, P^{0,+;q}_{2,2b}\}$ on the left-hand side 
and $\{F_{1,1},\, F_{1,2},\, F_{1,3},\, F_{1,4}\}\mapsto\{H_{2,1},\, H_{2,2},\, H_{2,3},\, H_{2,4}\}$ on the right-hand side;
\item[-] at twist 4, for $V^-$, we have the same relations~\eqref{rel1}-\eqref{rel4} with the replacement $\{S^{0,+;q}_{1,1a},\, S^{0,+;q}_{1,1b},\, P^{0,+;q}_{1,1a},\, P^{0,+;q}_{1,1b}\}\mapsto\{S^{0,+;q}_{3,1a},\, S^{0,+;q}_{3,1b},\, P^{0,+;q}_{3,1a},\, P^{0,+;q}_{3,1b}\}$ on the left-hand side and  $\{F_{1,1},\, F_{1,2},\, F_{1,3},\, F_{1,4}\}\mapsto\{F_{3,1},\, F_{3,2},\, F_{3,3},\, F_{3,4}\}$ on the right-hand side;
\end{itemize}

In the case of quark correlators involving operators with $\Delta S_z=0$ and $c_{\mathsf P}=-1$ (fourth column of table~\ref{Qstruct}):
\begin{itemize}
\item[-] at twist 2, for $A^+$, we have
\begin{align}
\label{rel1b}
S^{0,-;q}_{1,1a}&=\frac{1}{\sqrt{1-\xi^2}}\,G_{1,4},\\
\label{rel2b}
S^{0,-;q}_{1,1b}&=-\frac{1}{\sqrt{1-\xi^2}}\,G_{1,1},\\
\label{rel3b}
P^{0,-;q}_{1,1a}&=-\frac{1}{\sqrt{1-\xi^2}}\,\frac{\vec\Delta_T^2}{2M^2}\,G_{1,1}+\sqrt{1-\xi^2}\,G_{1,2},\\
\label{rel4b}
P^{0,-;q}_{1,1b}&=\frac{1}{\sqrt{1-\xi^2}}\,\frac{\vec k_T\cdot\vec \Delta_T}{2M^2}\,G_{1,1}+\sqrt{1-\xi^2}\,G_{1,3}-\frac{\xi}{2\sqrt{1-\xi^2}}\,G_{1,4};
\end{align}
\item[-] at twist 3, for $P$, we have the same relations~\eqref{rel1b}-\eqref{rel4b} with the replacement $\{S^{0,-;q}_{1,1a},\, S^{0,-;q}_{1,1b},\, P^{0,-;q}_{1,1a},\, P^{0,-;q}_{1,1b}\}\mapsto\{S^{0,-;q}_{2,2a},\, S^{0,-;q}_{2,2b},\, P^{0,-;q}_{2,2a},\, P^{0,-;q}_{2,2b}\}$ on the left-hand side and  $\{G_{1,1},\, G_{1,2},\, G_{1,3},\, G_{1,4}\}\mapsto\{E_{2,5},\, E_{2,6},\, E_{2,7},\, E_{2,8}\}$ on the right-hand side;
\item[-] at twist 3, for $T^{+-}$, we have the same relations~\eqref{rel1b}-\eqref{rel4b} with the replacement $\{S^{0,-;q}_{1,1a},\, S^{0,-;q}_{1,1b},\, P^{0,-;q}_{1,1a},\, P^{0,-;q}_{1,1b}\}\mapsto\{S^{0,-;q}_{2,1a},\, S^{0,-;q}_{2,1b},\, P^{0,-;q}_{2,1a},\, P^{0,-;q}_{2,1b}\}$ on the left-hand side and  $\{G_{1,1},\, G_{1,2},\, G_{1,3},\, G_{1,4}\}\mapsto\{H_{2,5},\, H_{2,6},\, H_{2,7},\, H_{2,8}\}$ on the right-hand side;
\item[-] at twist 4, for  $A^-$,  we have the same relations~\eqref{rel1b}-\eqref{rel4b} with the replacement $\{S^{0,-;q}_{1,1a},\, S^{0,-;q}_{1,1b},\, P^{0,-;q}_{1,1a},\, P^{0,-;q}_{1,1b}\}\mapsto\{S^{0,-;q}_{3,1a},\, S^{0,-;q}_{3,1b},\, P^{0,-;q}_{3,1a},\, P^{0,-;q}_{3,1b}\}$ on the left-hand side and  $\{G_{1,1},\, G_{1,2},\, G_{1,3},\, G_{1,4}\}\mapsto\{G_{3,1},\, G_{3,2},\, G_{3,3},\, G_{3,4}\}$ on the right-hand side.
\end{itemize}

The quark correlator at twist-3 with $V^{R(L)}$ is the only one with $\Delta S_z=\pm 1$ and $c_{\mathsf P}=+1$   (fifth column of table~\ref{Qstruct}). The relations between the two sets of GTMDs read
\begin{align}
S^{1,+;q}_{2,1a}&=\frac{1}{\sqrt{1-\xi^2}}\,\frac{\vec k_T\cdot\vec \Delta_T}{2M^2}
\left[F_{2,1}-2(1-\xi^2)\,F_{2,5}+\xi \,F_{2,7}\right]\nonumber\\
&\quad+\frac{1}{\sqrt{1-\xi^2}}\,\frac{\vec \Delta_T^2}{2M^2}\left[F_{2,2}-2(1-\xi^2)\,F_{2,6}+\xi \,F_{2,8}\right]\nonumber\\
&\quad-2 \sqrt{1-\xi^2}\,F_{2,3}-\sqrt{1-\xi^2}\,\frac{\vec k_T^2}{M^2}\,F_{2,4},
\label{rel1c}\\
\label{rel2c}
S^{1,+;q}_{2,1b}&=-\frac{1}{2\sqrt{1-\xi^2}}\left[F_{2,1}+2(1-\xi^2)\,F_{2,5}+\xi\, F_{2,7}\right],\\
\label{rel3c}
P^{1,+;q}_{2,1a}&=\frac{1}{\sqrt{1-\xi^2}}\,F_{2,1},\\
\label{rel4c}
P^{1,+;q}_{2,1b}&=\frac{1}{\sqrt{1-\xi^2}}\,F_{2,2},\\
\label{rel5c}
P'^{1,+;q}_{2,1a}&=-\frac{1}{\sqrt{1-\xi^2}}\,F_{2,7},\\
\label{rel6c}
P'^{1,+;q}_{2,1b}&=-\frac{1}{\sqrt{1-\xi^2}}\,F_{2,8},\\
D^{1,+;q}_{2,1a}&=\frac{1}{\sqrt{1-\xi^2}}\,\frac{\vec \Delta^2_T}{4\,\vec k_T\cdot\vec \Delta_T}\left[-F_{2,1}+2(1-\xi^2)\,F_{2,5}+\xi \,F_{2,7}\right]+\sqrt{1-\xi^2}\,F_{2,4},\label{rel7c}\\
D^{1,+;q}_{2,1b}&=\frac{1}{\sqrt{1-\xi^2}}\,\frac{\vec k^2_T}{4\,\vec k_T\cdot\vec \Delta_T}\left[-F_{2,1}+2(1-\xi^2)\,F_{2,5}+\xi \,F_{2,7}\right]\nonumber\\
&\quad+\frac{1}{2\sqrt{1-\xi^2}}\left[-F_{2,2}+2(1-\xi^2)\,F_{2,6}+\xi\, F_{2,8}\right].
\end{align}

In the case of  quark correlators  involving operators with $\Delta S_z=\pm 1$ and $c_{\mathsf P}=-1$ (last column of table~\ref{Qstruct}):
\begin{itemize}
\item[-] at twist 2, for $T^{R(L)+}$, we have
\begin{align}
S^{1,-;q}_{1,1a}&=\frac{1}{\sqrt{1-\xi^2}}\,\frac{\vec k_T\cdot\vec \Delta_T}{2M^2}\left[H_{1,1}+2(1-\xi^2)\, H_{1,5}-\xi \,H_{1,7}\right]\nonumber\\
&\quad+\frac{1}{\sqrt{1-\xi^2}}\,\frac{\vec \Delta_T^2}{2M^2}\left[H_{1,2}+2(1-\xi^2)\,H_{1,6}-\xi \,H_{1,8}\right]\nonumber\\
&\quad+2\sqrt{1-\xi^2}\,H_{1,3}+\sqrt{1-\xi^2}\,\frac{\vec k_T^2}{M^2}\,H_{1,4},\label{rel1d}\\
\label{rel2d}
S^{1,-;q}_{1,1b}&=\frac{1}{2\sqrt{1-\xi^2}}\left[-H_{1,1}+2(1-\xi^2)\,H_{1,5}+\xi\, H_{1,7}\right],\\
\label{rel3d}
P^{1,-;q}_{1,1a}&=\frac{1}{\sqrt{1-\xi^2}}\,H_{1,1},\\
\label{rel4d}
P^{1,-;q}_{1,1b}&=\frac{1}{\sqrt{1-\xi^2}}\,H_{1,2},\\
\label{rel5d}
P'^{1,-;q}_{1,1a}&=\frac{1}{\sqrt{1-\xi^2}}\,H_{1,7},\\
\label{rel6d}
P'^{1,-;q}_{1,1b}&=\frac{1}{\sqrt{1-\xi^2}}\,H_{1,8},\\
\label{rel7d}
D^{1,-;q}_{1,1a}&=\frac{1}{\sqrt{1-\xi^2}}\,\frac{\vec \Delta^2_T}{4\,\vec k_T\cdot\vec \Delta_T}\left[-H_{1,1}+2(1-\xi^2)\, H_{1,5}-\xi\, H_{1,7}\right]+\sqrt{1-\xi^2}\,H_{1,4},\\
D^{1,-;q}_{1,1b}&=\frac{1}{\sqrt{1-\xi^2}}\,\frac{\vec k^2_T}{4\,\vec k_T\cdot\vec \Delta_T}\left[-H_{1,1}+2(1-\xi^2)\,H_{1,5}-\xi \,H_{1,7}\right]\nonumber\\
&\quad+\frac{1}{2\sqrt{1-\xi^2}}\left[-H_{1,2}+2(1-\xi^2)\,H_{1,6}-\xi \,H_{1,8}\right].\label{rel8d}
\end{align}
\item[-] at twist 3, for $A^{R(L)}$, we have the same relations~\eqref{rel1d}-\eqref{rel8d} with the replacement $\{S^{1,-;q}_{1,1a},\, S^{1,-;q}_{1,1b},\, P^{1,-;q}_{1,1a},\, P^{1,-;q}_{1,1b},\, P'^{1,-;q}_{1,1a},\,P'^{1,-;q}_{1,1b},\, D^{1,-;q}_{1,1a},\, D^{1,-;q}_{1,1b}\}$\\$\mapsto\{S^{1,-;q}_{2,1a},\, S^{1,-;q}_{2,1b},\, P^{1,-;q}_{2,1a},\, P^{1,-;q}_{2,1b},\, P'^{1,-;q}_{2,1a},\, P'^{1,-;q}_{2,1b},\, D^{1,-;q}_{2,1a},\, D^{1,-;q}_{2,1b}\}$ on the left-hand side and $\{H_{1,1},\, H_{1,2},\, H_{1,3},\, H_{1,4},\, H_{1,5},\, H_{1,6},\, H_{1,7},\, H_{1,8}\}$\\$\mapsto \{G_{2,1},\, G_{2,2},\, G_{2,3},\, G_{2,4},\, G_{2,5},\, G_{2,6},\, G_{2,7},\, G_{2,8}\}$ on the right-hand side;
\item[-] at twist 4, for $T^{R(L)-}$, we have the same relations~\eqref{rel1d}-\eqref{rel8d} with the replacement $\{S^{1,-;q}_{1,1a},\, S^{1,-;q}_{1,1b},\, P^{1,-;q}_{1,1a},\, P^{1,-;q}_{1,1b},\, P'^{1,-;q}_{1,1a},\, P'^{1,-;q}_{1,1b},\, D^{1,-;q}_{1,1a},\, D^{1,-;q}_{1,1b}\}$\\$\mapsto\{S^{1,-;q}_{3,1a},\, S^{1,-;q}_{3,1b},\, P^{1,-;q}_{3,1a},\, P^{1,-;q}_{3,1b},\, P'^{1,-;q}_{3,1a},\, P'^{1,-;q}_{3,1b},\, D^{1,-;q}_{3,1a},\, D^{1,-;q}_{3,1b}\}$ on the left-hand side and $\{H_{1,1},\, H_{1,2},\, H_{1,3},\, H_{1,4},\, H_{1,5},\, H_{1,6},\, H_{1,7},\, H_{1,8}\}$\\$\mapsto\{H_{3,1},\, H_{3,2},\, H_{3,3},\, H_{3,4},\, H_{3,5},\, H_{3,6},\, H_{3,7},\, H_{3,8}\}$ on the right-hand side.
\end{itemize}



\begin{thebibliography}{99}

\bibitem{Mueller:1998fv} 
  D.~Mueller, D.~Robaschik, B.~Geyer, F.~M.~Dittes and J.~Horejsi,
  {\it Wave functions, evolution equations and evolution kernels from light ray operators of QCD,}
{\em   Fortsch.\ Phys}.\  {\bf 42}, 101 (1994)
   [\href{http://arxiv.org/abs/hep-ph/9812448}{{\tt hep-ph/9812448}}].

\bibitem{Ji:1996ek} 
  X.~-D.~Ji,
 {\it Gauge invariant decomposition of nucleon spin and its spin - off,}
 {\em  Phys.\ Rev.\ Lett.}\  {\bf 78}, 610 (1997)
   [\href{http://arxiv.org/abs/hep-ph/9603249}{{\tt hep-ph/9603249}}].

\bibitem{Radyushkin:1996nd} 
  A.~V.~Radyushkin,
  {\it Scaling limit of deeply virtual Compton scattering,}
  {\em Phys.\ Lett.} {\bf B 380}, 417 (1996) 
  [\href{http://arxiv.org/abs/hep-ph/9604317}{{\tt hep-ph/9604317}}].  

\bibitem{Belitsky:2005qn}
  A.~V.~Belitsky and A.~V.~Radyushkin,
  {\it Unraveling hadron structure with generalized parton distributions},
  {\em Phys.\ Rept.}  {\bf 418}, 1 (2005) 
  [\href{http://arxiv.org/abs/hep-ph/0504030}{{\tt hep-ph/0504030}}].



\bibitem{Goeke:2001tz} 
  K.~Goeke, M.~V.~Polyakov and M.~Vanderhaeghen,
  {\it Hard exclusive reactions and the structure of hadrons,}
  {\em Prog.\ Part.\ Nucl.\ Phys.}\  {\bf 47}, 401 (2001)
    [\href{http://arxiv.org/abs/hep-ph/0106012}{{\tt hep-ph/0106012}}].  
  
\bibitem{Diehl:2003ny}
  M.~Diehl,
  {\it Generalized parton distributions,}
{\em  Phys.\ Rept.} {\bf 388}, 41 (2003) 
 [\href{http://arxiv.org/abs/hep-ph/0307382}{{\tt hep-ph/0307382}}].
  
\bibitem{Boffi:2007yc}
  S.~Boffi and B.~Pasquini,
  {\it Generalized parton distributions and the structure of the nucleon},
  {\em Riv.\ Nuovo Cim.}  {\bf 30}, 387 (2007) 
  [\href{http://arxiv.org/abs/0711.2625}{{\tt arXiv:0711.2625}}].


\bibitem{Mulders:1995dh} 
  P.~J.~Mulders and R.~D.~Tangerman,
  {\it The Complete tree level result up to order 1/Q for polarized deep inelastic leptoproduction,}
 {\em Nucl.\ Phys.}\  {\bf B 461}, 197 (1996)
  [{\em Erratum-ibid.}{\bf\ B  484}, 538 (1997)]
   [\href{http://arxiv.org/abs/hep-ph/9510301}{{\tt hep-ph/9510301}}]. 

\bibitem{Barone:2001sp} 
  V.~Barone, A.~Drago and P.~G.~Ratcliffe,
  {\it Transverse polarisation of quarks in hadrons,}
 {\em Phys.\ Rept.} {\bf 359}, 1 (2002)
 [\href{http://arxiv.org/abs/hep-ph/0104283}{{\tt hep-ph/0104283}}].   

\bibitem{Bacchetta:2006tn} 
  A.~Bacchetta, M.~Diehl, K.~Goeke, A.~Metz, P.~J.~Mulders and M.~Schlegel,
  {\it Semi-inclusive deep inelastic scattering at small transverse momentum,}
 {\em JHEP} {\bf 0702}, 093 (2007)
 [\href{http://arxiv.org/abs/hep-ph/0611265}{{\tt hep-ph/0611265}}].   

\bibitem{D'Alesio:2007jt} 
  U.~D'Alesio and F.~Murgia,
  {\it Azimuthal and Single Spin Asymmetries in Hard Scattering Processes,}
 {\em  Prog.\ Part.\ Nucl.\ Phys.}  {\bf 61}, 394 (2008)
  [\href{http://arxiv.org/abs/0712.4328}{{\tt arXiv:0712.4328}}].  

\bibitem{Boer:2011fh} 
  D.~Boer, M.~Diehl, R.~Milner, R.~Venugopalan, W.~Vogelsang, D.~Kaplan, H.~Montgomery and S.~Vigdor {\it et al.},
  {\it Gluons and the quark sea at high energies: Distributions, polarization, tomography,}
 [\href{http://arxiv.org/abs/arXiv:1108.1713}{{\tt arXiv:1108.1713}}].

\bibitem{Burkardt:2002ks} 
  M.~Burkardt,
  {\it Impact parameter dependent parton distributions and transverse single spin asymmetries,}
  {\em Phys.\ Rev.} {\bf D 66}, 114005 (2002)
 [\href{http://arxiv.org/abs/hep-ph/0209179}{{\tt hep-ph/0209179}}].   

\bibitem{Burkardt:2003uw}
  M.~Burkardt,
  {\it Chromodynamic lensing and transverse single spin asymmetries},
  {\em Nucl.\ Phys.} {\bf A 735}, 185 (2004) 
  [\href{http://arxiv.org/abs/hep-ph/0302144}{{\tt hep-ph/0302144}}].

\bibitem{Burkardt:2003je}
  M.~Burkardt and D.~S.~Hwang,
  {\it Sivers asymmetry and generalized parton distributions in impact  parameter
  space},
  {\em Phys.\ Rev.}   {\bf D 69}, 074032 (2004) 
  [\href{http://arxiv.org/abs/hep-ph/0309072}{{\tt hep-ph/0309072}}].


\bibitem{Diehl:2005jf}
  M.~Diehl and Ph.~H\"agler,
  {\it Spin densities in the transverse plane and generalized transversity
  distributions},
  {\em Eur.\ Phys.\ J.}   {\bf C 44}, 87 (2005) 
  [\href{http://arxiv.org/abs/hep-ph/0504175}{{\tt hep-ph/0504175}}].


\bibitem{Burkardt:2005hp} 
  M.~Burkardt,
  {\it Transverse deformation of parton distributions and transversity decomposition of angular momentum,}
  {\em Phys.\ Rev.}\  {\bf D 72}, 094020 (2005)
  [\href{http://arxiv.org/abs/hep-ph/0505189}{{\tt hep-ph/0505189}}].

\bibitem{Meissner:2007rx}
  S.~Meissner, A.~Metz and K.~Goeke,
  {\it Relations between generalized and transverse momentum dependent parton
  distributions},
  {\em Phys.\ Rev.} {\bf D 76}, 034002 (2007) 
  [\href{http://arxiv.org/abs/hep-ph/0703176}{{\tt hep-ph/0703176}}].

\bibitem{Lu:2006kt} 
  Z.~Lu and I.~Schmidt,
  {\it Connection between the Sivers function and the anomalous magnetic moment,}
  {\em Phys.\ Rev.}\  {\bf D 75}, 073008 (2007)
  [\href{http://arxiv.org/abs/hep-ph/0611158}{{\tt hep-ph/0611158}}].

\bibitem{Meissner:2009ww}
  S.~Meissner, A.~Metz and M.~Schlegel,
  {\it Generalized parton correlation functions for a spin-1/2 hadron},
  {\em JHEP} {\bf 0908}, 056 (2009) 
  [\href{http://arxiv.org/abs/0906.5323}{{\tt arXiv:0906.5323}}].

\bibitem{Meissner:2008ay}
  S.~Meissner, A.~Metz, M.~Schlegel and K.~Goeke,
  {\it Generalized parton correlation functions for a spin-0 hadron},
  {\em JHEP} {\bf 0808}, 038 (2008) 
    [\href{http://arxiv.org/abs/0805.3165}{{\tt arXiv:0805.3165}}].
\bibitem{Collins:2007ph} 
  J.~C.~Collins, T.~C.~Rogers and A.~M.~Stasto,
  {\it Fully unintegrated parton correlation functions and factorization in lowest-order hard scattering,}
  {\em Phys.\ Rev.}\  {\bf D 77}, 085009 (2008)
 [\href{http://arxiv.org/abs/arXiv:0708.2833}{{\tt arXiv:0708.2833}}].

\bibitem{Rogers:2008jk} 
  T.~C.~Rogers,
  {\it Next-to-Leading Order Hard Scattering Using Fully Unintegrated Parton Distribution Functions,}
  {\em Phys.\ Rev.}\  {\bf D 78}, 074018 (2008)
 [\href{http://arxiv.org/abs/arXiv:0807.2430}{{\tt arXiv:0807.2430}}].

\bibitem{Vanderhaeghen:1999xj} 
  M.~Vanderhaeghen, P.~A.~M.~Guichon and M.~Guidal,
  {\it Deeply virtual electroproduction of photons and mesons on the nucleon: Leading order amplitudes and power corrections,}
  {\em Phys.\ Rev.}\  {\bf D 60}, 094017 (1999)
  [\href{http://arxiv.org/abs/hep-ph/9905372}{{\tt hep-ph/9905372}}].

\bibitem{Diehl:2007hd} 
  M.~Diehl and W.~Kugler,
  {\it Next-to-leading order corrections in exclusive meson production,}
  {\em Eur.\ Phys.\ J.} \  {\bf C 52}, 933 (2007)
 [\href{http://arxiv.org/abs/arXiv:0708.1121}{{\tt arXiv:0708.1121}}].

\bibitem{Goloskokov:2007nt} 
  S.~V.~Goloskokov and P.~Kroll,
  {\it The Role of the quark and gluon GPDs in hard vector-meson electroproduction,}
  {\em Eur.\ Phys.\ J.} \  {\bf C 53}, 367 (2008)
 [\href{http://arxiv.org/abs/arXiv:0708.3569}{{\tt arXiv:0708.3569}}].


\bibitem{Martin:1999wb} 
  A.~D.~Martin, M.~G.~Ryskin and T.~Teubner,
  {\it $Q^2$ dependence of diffractive vector meson electroproduction,}
  {\em Phys.\ Rev.} \  {\bf D 62}, 014022 (2000)
  [\href{http://arxiv.org/abs/hep-ph/9912551}{{\tt hep-ph/9912551}}].

\bibitem{Khoze:2000cy} 
  V.~A.~Khoze, A.~D.~Martin and M.~G.~Ryskin,
  {\it Can the Higgs be seen in rapidity gap events at the Tevatron or the LHC?,}
  {\em Eur.\ Phys.\ J.} \  {\bf C 14}, 525 (2000)
  [\href{http://arxiv.org/abs/hep-ph/0002072}{{\tt hep-ph/0002072}}].

\bibitem{Albrow:2008pn} 
  M.~G.~Albrow {\it et al.}  [FP420 R and D Collaboration],
  {\it The FP420 \&  Project: Higgs and New Physics with forward protons at the LHC,}
  {\em JINST} {\bf 4}, T10001 (2009)
 [\href{http://arxiv.org/abs/arXiv:0806.0302}{{\tt arXiv:0806.0302}}].

\bibitem{Martin:2009ku} 
  A.~D.~Martin, M.~G.~Ryskin and V.~A.~Khoze,
 {\it Forward Physics at the LHC,}
  {\em Acta Phys.\ Polon.} \  {\bf B 40}, 1841 (2009)
 [\href{http://arxiv.org/abs/arXiv:0903.2980}{{\tt arXiv:0903.2980}}].

\bibitem{Martin:2001ms} 
  A.~D.~Martin and M.~G.~Ryskin,
  {\it Unintegrated generalized parton distributions,}
  {\em Phys.\ Rev.} \  {\bf D 64}, 094017 (2001)
  [\href{http://arxiv.org/abs/hep-ph/0107149}{{\tt hep-ph/0107149}}].


\bibitem{Ji:2003ak}
  X.~d.~Ji,
  {\it Viewing the proton through ``color''-filters},
  {\em Phys.\ Rev.\ Lett.}  {\bf 91}, 062001 (2003) 
    [\href{http://arxiv.org/abs/hep-ph/0304037}{{\tt hep-ph/0304037}}].

\bibitem{Belitsky:2003nz}
  A.~V.~Belitsky, X.~d.~Ji and F.~Yuan,
  {\it Quark imaging in the proton via quantum phase-space distributions},
  {\em Phys.\ Rev.} {\bf D 69}, 074014 (2004) 
    [\href{http://arxiv.org/abs/hep-ph/0307383}{{\tt hep-ph/0307383}}].

\bibitem{Lorce:2011dv}
  C.~Lorc\'e, B.~Pasquini and M.~Vanderhaeghen,
 {\it Unified framework for generalized and transverse-momentum dependent parton
  distributions within a 3Q light-cone picture of the nucleon,}
  {\em JHEP} {\bf 1105}, 041 (2011)  
 [\href{http://arxiv.org/abs/1102.4704}{{\tt arXiv:1102.4704}}].

\bibitem{Lorce:2011kd} 
  C.~Lorc\'e and B.~Pasquini,
{\it Quark Wigner Distributions and Orbital Angular Momentum,}
  {\em Phys. Rev.}  {\bf D 84}, 014015 (2011)
 [\href{http://arxiv.org/abs/arXiv:1106.0139}{{\tt arXiv:1106.0139}}].    
 
 
  
\bibitem{Lorce:2011ni} 
  C.~Lorc\'e, B.~Pasquini, X.~Xiong and F.~Yuan,
  {\it The quark orbital angular momentum from Wigner distributions and light-cone wave functions,}
  {\em Phys.\ Rev.}  {\bf  D 85}, 114006 (2012)  
 [\href{http://arxiv.org/abs/arXiv:1111.4827}{{\tt arXiv:1111.4827}}].

\bibitem{Hatta:2011ku} 
  Y.~Hatta,
  {\it Notes on the orbital angular momentum of quarks in the nucleon,}
  {\em Phys.\ Lett.}  {\bf B 708}, 186 (2012)
 [\href{http://arxiv.org/abs/arXiv:1111.3547}{{\tt arXiv:1111.3547}}].

\bibitem{Ji:2012sj} 
  X.~Ji, X.~Xiong and F.~Yuan,
  {\it Proton Spin Structure from Measurable Parton Distributions,}
  {\em Phys.\ Rev.\ Lett.}  {\bf 109}, 152005 (2012)
 [\href{http://arxiv.org/abs/arXiv:1202.2843}{{\tt arXiv:1202.2843}}].

\bibitem{Pijlman:2006vm}
  F.~Pijlman,
  {\it Single spin asymmetries and gauge invariance in hard scattering processes,}
   [\href{http://arxiv.org/abs/hep-ph/0604226}{{\tt hep-ph/0604226}}].
 

\bibitem{Buffing:2011mj} 
  M.~G.~A.~Buffing and P.~J.~Mulders,
  {\it Gauge links for transverse momentum dependent correlators at tree-level,}
  {\em JHEP} {\bf 1107}, 065 (2011)
 [\href{http://arxiv.org/abs/arXiv:1105.4804}{{\tt arXiv:1105.4804}}].

\bibitem{Buffing:2012sz} 
  M.~G.~A.~Buffing, A.~Mukherjee and P.~J.~Mulders,
  {\it Generalized Universality of Higher Transverse Moments of Quark TMD Correlators,}
  {\em Phys.\ Rev.} \  {\bf D 86}, 074030 (2012)
 [\href{http://arxiv.org/abs/arXiv:1207.3221}{{\tt arXiv:1207.3221}}].

\bibitem{Buffing:2013kca} 
  M.~G.~A.~Buffing, A.~Mukherjee and P.~J.~Mulders,
 {\it Generalized Universality of Definite Rank Gluon Transverse Momentum Dependent Correlators,}
[\href{http://arxiv.org/abs/arXiv:1306.5897}{{\tt arXiv:1306.5897}}].

\bibitem{Brodsky:2006ez} 
  S.~J.~Brodsky, S.~Gardner and D.~S.~Hwang,
  {\it Discrete symmetries on the light front and a general relation connecting nucleon electric dipole and anomalous magnetic moments},
  {\em Phys.\ Rev.} {\bf D 73},  036007  (2006) 
   [\href{http://arxiv.org/abs/hep-ph/0601037}{{\tt hep-ph/0601037}}].
 
\bibitem{Soper:1972xc} 
  D.~E.~Soper,
 {\it Infinite-momentum helicity states,}
 {\em 
  Phys.\ Rev.}  {\bf D 5}, 1956 (1972).
  
\bibitem{Carlson:2003je} 
  C.~E.~Carlson and C.~-R.~Ji,
  {\it Angular conditions, relations between Breit and light front frames, and subleading power corrections,}
  {\em 
  Phys.\ Rev.}  {\bf D 67}, 116002 (2003) 
  [\href{http://arxiv.org/abs/hep-ph/0301213}{{\tt hep-ph/0301213}}].  

\bibitem{Diehl:2001pm}
  M.~Diehl,
  {\it Generalized parton distributions with helicity flip},
  {\em Eur.\ Phys.\ J.} {\bf C 19}, 485 (2001) 
  [\href{http://arxiv.org/abs/hep-ph/0101335}{{\tt hep-ph/0101335}}].

\bibitem{Goeke:2005hb}
  K.~Goeke, A.~Metz and M.~Schlegel,
  {\it Parameterization of the quark-quark correlator of a spin-1/2 hadron,}
 {\em   Phys.\ Lett.} {\bf B 618}, 90 (2005)
 [\href{http://arxiv.org/abs/hep-ph/0504130} {{\tt hep-ph/0504130}}].

\bibitem{Avakian:2010br}
  H.~Avakian, A.~V.~Efremov, P.~Schweitzer and F.~Yuan,
  {\it The transverse momentum dependent distribution functions in the bag
  model},
  {\em Phys.\ Rev.} {\bf D 81}, 074035 (2010) 
  [\href{http://arxiv.org/abs/1001.5467}{{\tt arXiv:1001.5467}}].

\bibitem{Mulders:2000sh} 
  P.~J.~Mulders and J.~Rodrigues,
  {\it Transverse momentum dependence in gluon distribution and fragmentation functions,}
  {\em 
  Phys.\ Rev.} {\bf D 63}, 094021  (2001) 
 [\href{http://arxiv.org/abs/hep-ph/0009343}{{\tt hep-ph/0009343}}].    
  
  \bibitem{Kiptily:2002nx}
  D.~V.~Kiptily and M.~V.~Polyakov,
  {\it Genuine twist three contributions to the generalized parton distributions from instantons,}
  {\em 
  Eur.\ Phys.\ J.} {\bf C 37},105 (2004)
   [\href{http://arxiv.org/abs/hep-ph/0212372}{{\tt hep-ph/0212372}}].
     

 
\end{thebibliography}
\end{document}